%% file: smc8.tex
\input macro

\magnification=\magstep1
\baselineskip=11pt
\def\ni{\noindent}
\def\r{r_\alpha}
\def\d{d_\alpha}

\def\dr{\delta r_{\alpha}}
\def\om{\Omega_o}
\def\os{\hat \Omega_*}
\def\omp{\Omega_p}

\def\Dk{D_k^{(2 m)}}

\def\Ar{A^{(r)}_{\alpha}}
\def\g{\Gamma_\alpha}
\def\frac#1#2{{#1\over #2}}
\def\kap{\kappa_{\alpha \beta \beta}}
\def\refind{\noindent \hangindent=2pc \hangafter=1}

\centerline{ }
\vskip 0.5truecm
\centerline{\bf On the orbital decay of the PSR J0045-7319 Binary}
\vskip 0.7truecm
\centerline{Pawan Kumar$^{*}$}
\centerline{Institute for Advanced Study, Princeton, NJ 08540}
\medskip
\centerline{and}
\medskip
\centerline{Eliot J. Quataert}
\centerline{Harvard-Smithsonian Center for Astrophysics, 60 Garden St., 
Cambridge, MA 02138}

\medskip\bigskip
\centerline{\bf ABSTRACT}
\bigskip

Recent observations of PSR J0045-7319, a radio pulsar in a close eccentric 
orbit with a massive main sequence B-star companion, indicate that the 
system's orbital period is decreasing on a timescale $\sim 5$x$10^{5}$ years 
(Kaspi et al. 1996).  Timing observations of PSR J0045-7319 also indicate that 
the B-star is  rotating rapidly, perhaps close to its breakup rotation rate 
(Lai et al. 1995, Kaspi et al. 1996). For rapid (super-synchronous) prograde 
rotation of the B-star, tidal dissipation leads to an increasing orbital 
period for the binary system, while for retrograde rotation of any magnitude, 
the orbital period decreases with time. We show that if tidal effects are to
account for the observed orbital decay of the PSR J0045-7319 binary, the 
B-star must have retrograde rotation. This implies that the supernova that 
produced the pulsar in this binary system likely had a dipole anisotropy.

For a reasonably wide range of retrograde rotation rates, the 
energy in the dynamical tide of the B-star needs to be dissipated in about one
orbital period in order to account for the observed orbital evolution
time for the PSR J0045-7319 binary. We show, however, that the
radiative dissipation of the dynamical tide in a rigidly rotating
B-star is too inefficient by a factor of $\approx$ 10$^3$, regardless of the
magnitude of the rotation rate. We describe how, when the surface of
the B-star is rotating nearly synchronously (which is expected
from the work of Goldreich and Nicholson, 1989), the energy in the
dynamical tide is dissipated in less than an orbital period, thus
reconciling the theoretical and observed rates of orbital evolution.

Nonlinear parametric decay of the equilibrium tide, for rigid
retrograde rotation of the B-star, may also be able to explain the
observed rate of orbital evolution, though the margin of instability
is too small to draw definitive conclusions about the relevance of
this process for the PSR J0045-7319 Binary.

\ni{\it Subject headings:} stars: binaries --- stars: oscillations ---
stars: rotation --- stars: early-type

~~~~~~~~~~~~~~~~~~pulsars: individual (PSR J0045-7319)

\vskip .8truecm
\ni$^*$Alfred P. Sloan Fellow \& NSF Young Investigator

\noindent email addresses: pk@sns.ias.edu (P. Kumar)

\hskip 55pt equataert@cfa.harvard.edu (E.J. Quataert)

\vfill\eject
\baselineskip=13pt
\centerline{\bf \S1. Introduction}

The recent discovery of two radio pulsars in binary systems with main
sequence star companions has provided the opportunity to study tidal
interactions with unprecedented precision (McConnell et al. 1991,
Kaspi et al. 1994; Johnston et al. 1992). Of these two radio pulsars, 
the one in the Small Magellanic Cloud (SMC), PSR J0045-7319 
(which has a spin period of 0.93 s), is particularly 
interesting for studying tidal interactions because the periastron separation
between the neutron star and its companion, a main sequence B-star of
mass $\approx 8.8 M_\odot$, is only about 4 times the B-star radius.
The monitoring of pulse arrival times for the SMC pulsar has enabled
accurate determination of several orbital parameters such as the
period (51 days), eccentricity (0.808) and their time derivatives.
Kaspi et al. (1996)
report that the orbital period for the PSR J0045-7319 Binary is
decreasing on a time scale of $\sim 5 \times 10^5$ years and that the
orbital eccentricity is probably evolving more slowly.  All relevant
parameters for this system are included in table 1 for easy access.
The work of Lai et al. (1995), Kaspi et al. (1996) and Bell et al. (1995) 
has also provided evidence that the B-star companion is rotating 
close to its break-up rotation rate and that its spin axis is misaligned with 
the angular momentum of the system.

The evolution time for the SMC Binary's orbit due to the radiative
dissipation of the dynamical tide (in a nonrotating B-star) is $\sim
10^9$ years (see \S2).  It has been suggested (Lai, 1996), however, that the
observed rapid evolution of the SMC Binary can be accounted for by the
radiative dissipation of the dynamical tide, provided that the B-star
companion to PSR J0045-7319 has a rapid retrograde rotation rate of
~$\hat \Omega_* \approx -0.4$, where $\hat \Omega_*$ is the component
of the stellar rotation rate perpendicular to the orbital plane in
units of $(GM_*/R_*^3)^{1/2}$. Retrograde rotation has the obvious
advantage that the frequency of the tide seen from the rest frame of
the B-star is considerably higher than for prograde rotation, causing
excitation of very low order g-modes to large amplitudes.  However,
the dissipation times for low order g-modes, which set the rate at
which orbital energy is extracted by the star, are considerably longer
than the dissipation times of the higher order g-modes excited in a
non-rotating star. This offsets the advantage of having increased mode
amplitudes in a retrograde rotating star. For a concise and physically
motivated derivation of this result please see Kumar \& Quataert (1996).

In the next section we describe the orbital evolution caused by the
radiative damping of the dynamical tide in a uniformly rotating star,
and discuss the probability that mode resonances may substantially
speed up the rate of orbital evolution.  In \S3 we consider several
alternate dissipation mechanisms, in particular nonlinear parametric
instability (Kumar \& Goodman 1996) and enhanced radiative dissipation
in a differentially rotating star, in order to assess if they might be
important for understanding the orbital evolution of the PSR
J0045-7319 Binary. Please note that all frequencies in this paper,
including rotational and orbital angular speeds, are given in micro-Hz 
($\mu$Hz) unless explicitly stated otherwise.

\bigskip
\centerline{\bf \S2 Dynamical tide in a uniformly rotating star}
\medskip

Consider a binary system where the primary star has mass $M_*$, and
its companion, the secondary star, has mass $M$. The primary is
assumed to be rotating as a solid body with the component of its
angular velocity normal to the orbital plane of $\Omega_*$.

The tidal forcing of a star predominantly excites gravity modes
(g-modes) since they have periods comparable to the period of the
tidal forcing.  The g-modes in a uniformly rotating star are specified
uniquely by three numbers: $n$ the number of radial nodes, $\lambda$,
which is in general not an integer, is a generalization of the
spherical harmonic degree $\ell$, and $m$, the azimuthal
order.\footnote{$^1$}{Because the modification to the mode structure
in the rotating star is not central to our analysis, we have relegated
a discussion of it to Appendix B; in addition, we often use $\ell$
in place of $\lambda$.} The mode amplitude in the rotating
reference frame of the star, $A_\alpha^{(r)}$ ($\alpha$ denotes the collective
index $n,\lambda,m$), for the dominant quadrupole tide, can be calculated 
by decomposing the tidal forcing function in terms of its Fourier 
coefficients and is given below (see Kumar et al. 1995, and Quataert et al.
1996 for details)

\eqnam{\ampli}
$$ \Ar(t)  \approx f_\alpha \exp(im\Omega_* t)
\sum_{k=1}^\infty {\Dk\exp\left[\mp ik\Omega_0 t \mp i\phi_k\right]
\over \sqrt{ [{\r^2} - (k-|m|s)^2]^2 + \d^2(k - |m|s)^2}} 
 \eqno(\new)$$
\ni where
$$ f_\alpha = \frac{4 \pi}{5} \frac{G M \omega^2_\alpha
Q_\alpha}{a^3},\quad \r = \frac{\omega_\alpha}{\om}, \quad \d =
\frac{\g}{\om}, \quad s = {\Omega_*\over\om},\quad \tan\phi_k =
{\d(|m|s-k)\over {\r^2} - (k - |m|s)^2}, \eqno(\new)$$ 
$\om=\sqrt{G (M+M_*)/a^3}$ is the
orbital frequency, $a$ is the semi-major axis of the orbit, $\Dk$ are
the fourier coefficients of the quadrupole tidal forcing function,
$\g$ is the mode's energy dissipation rate, and $Q_\alpha$ is the
overlap of the mode's normalized Eulerian density perturbation,
$\delta \rho_\alpha$, with the quadrupole tidal potential.  The
negative sign in the exponent of equation (\ampli) is considered when
$m$ is positive.

We normalize our mode eigenfunctions so that the mode energy is equal to
the square of the mode amplitude. Thus,
the energy in mode $\alpha$ due to tidal forcing, as seen by an
observer corotating with the star, is obtained by taking the square of
the resonant term in the Fourier series, {\it i.e.},

\eqnam{\energy}
$$E^{(r)}_\alpha \approx {f_\alpha^2 \left|D_{k_\alpha}^{(\ell m)} \right|^2
\over \left[ r_\alpha^2 - (k_\alpha - |m|s)^2\right]^2 + \d^2(k_\alpha
- |m|s)^2}, \eqno(\new)$$ 
where $k_\alpha$ is a positive integer that minimizes the denominator,
{\it i.e.}, $k_\alpha = \lfloor s |m| \pm
r_\alpha \rfloor$, where $\lfloor x \rfloor$ denotes the nearest
integer to $x$.  The net energy in the dynamical tide is $E^{(r)}_{tide} =
\sum_\alpha E^{(r)}_\alpha$.

It is easy to obtain $\Dk$ numerically, and, for $m=\pm 2$, it peaks at $k
\approx 2\Omega_p/\Omega_o$, where $\omp = (1-e)^{-3/2}(1+e)^{1/2} \om$ is 
the orbital angular speed of the star at periastron and $e$ is the
orbital eccentricity. The peak value of $D_k^{(22)}$ is $\approx
(1+e)^{-1}\sqrt{15/128\pi} (\omp/\om)$ while for $k \gta
5\Omega_p/\Omega_o$, $D_k^{(22)} \approx \exp(-1.3 k\om/\omp)/(1-e)^{3/2}$.
The $m = 0$ Fourier coefficients, $D_k^{(20)}$, are roughly constant
for $k$ up to $2\Omega_p/\Omega_o$ and decrease exponentially at larger $k$.
Moreover, the $D_k^{(20)}$ are
typically smaller than the $D_k^{(2\pm2)}$ by a factor of at least a few and so
most of the tidal energy resides in the $m = \pm 2$ modes unless 
$\Omega_*\approx \omp$. 
The peak of $D_k^{(22)}$ at $k \approx 2\omp/\om$ and the exponential fall
off at higher harmonics implies that the mode with the largest energy
has a frequency $\omega_\alpha \approx \Omega_{tide}$ (measured in 
a  frame corotating with the star), where $\Omega_{tide} \equiv 2|\omp - \Omega_*|$ 
is twice the angular speed of the secondary at periastron, as seen in the rotating
reference frame of the B-star. Higher frequency modes, while they have
larger $Q_\alpha$, have less energy because $\Ar$ decreases
exponentially for larger frequencies.  Lower frequency modes have less
energy because they have a larger number of radial nodes and thus
smaller $Q_\alpha$.

The effect of stellar rotation, for $0 < \Omega_* <
\omp$, is to decrease the energy in modes since the resonance is shifted
to higher harmonics. The modes that are most efficiently excited have
$\omega_\alpha\approx \Omega_{tide}$, and so $E^{(r)}_{tide}$ decreases
with increasing $\Omega_*$ as lower frequency g-modes with smaller
wavelengths and smaller $Q_\alpha$ are excited.  For $\Omega_*
\approx \omp$ or $\Omega_{tide} \approx 0$, the energy in the dynamical
tide reaches its minimum value.\footnote{$^2$}{This corresponds to the $m = 0$ and $m = \pm 2$ modes being excited to comparable amplitudes.
For $\Omega_{tide} \ne 0$, the $m = \pm 2$ modes are excited to
greater amplitudes than the $m = 0$ modes. }

For $\Omega_* > \omp$, however, $\Omega_{tide}$ increases with
increasing $\Omega_*$, higher frequency modes are excited by the tidal
forcing and so $E^{(r)}_{tide}$ increases. It is also clear from equation
(\last) that for retrograde rotation ($\Omega_* < 0$) the tidal
frequency is larger and the energy in the dynamical tide is greater
(as was pointed out by Lai, 1996).  All of these effects are clearly
seen in Figure (1a), which is a plot of $E^{(r)}_{tide}$, the sum of
the energy in all modes (including $m=0$ \& m = $\pm2$), as seen by an 
observer corotating with the B-star, 
for the SMC binary system as a function of the B-star's 
rotation rate.

The energy in a mode is proportional to $\dr^{-2}$,
where $\dr\equiv r_\alpha - |k_\alpha- |m|s|$ (see eq.  [\last]) and
so close resonances can significantly enhance the energy in the
dynamical tide.  The effect of resonances is evident in Figure (1).

The mode energy as seen by an observer corotating with the star,
$E^{(r)}_\alpha$, does not include the work done by the tidal force on
the velocity field associated with the rotation of the star, which
changes the star's rotational energy. The mode energy as seen by an
inertial observer includes this contribution and is given by
$E^{(i)}_{\alpha} = E^{(r)}_\alpha + \Omega_* L_\alpha$, where
$L_\alpha$, the angular momentum associated with a gravity wave of
frequency $\omega_\alpha$ and azimuthal order $m$ can be shown to be
equal to $-m E_\alpha/{\omega_\alpha}$. We note that
$E_\alpha$ and $\omega_\alpha$ are frame dependent quantities, but the
ratio $E_\alpha/\omega_\alpha \propto L_\alpha$ is frame
independent. The net tidal energy as seen by an inertial observer,
$E^{(i)}_{tide} = \sum_\alpha E^{(i)}_\alpha$, is shown in Figure (1b)
for the SMC binary system (we note that the results presented in fig. 1 were
calculated using the entire fourier series expansion
of $E^{(i)}_{tide}$ and $E^{(r)}_{tide}$, rather than just the 
resonant term; see Appendix A).

Since (for $m \ne 0$) the azimuthal propagation of tidally excited waves, 
as seen by an inertial observer, is always in the direction of the orbital
motion, the sign of $m/\omega_\alpha$ is negative and is independent
of $\Omega_*$. This implies that $L_\alpha$ and $E^{(i)}_\alpha$ have
the same sign.  Since the angular momentum deposited in the star is
positive for $\Omega_*\lta\Omega_p$, the tidal energy deposited in the
star is also positive, which leads to a decrease in the orbital period
with time. For $\Omega_*\gta\Omega_p$, on the other hand, the sign of
the angular momentum deposited reverses and so there is a net transfer
of energy from the star to the orbit. Physically, for $\Omega_* \gta
\Omega_p$, the energy taken out of the spin of the star (which is
being slowed down) exceeds the energy deposited in modes and so the
orbital energy and period of the binary system increase with time.  A
more detailed discussion of these results is given in Appendix A.

In order to calculate the orbital evolution we need to know the
dissipation rate of the dynamical tide, which we calculate in the next
section.

\bigskip
\centerline{\bf 2a. Radiative damping of g-modes}
\medskip

The radiative damping times for the low order quadrupole g-modes of a
8.8 M$_\odot$, slightly evolved, main sequence star are given in Table 2 
(the model was kindly provided by the Yale group). The dissipation was 
calculated by solving the fully non-adiabatic oscillation equations (see 
Unno et al. 1989 or Kumar 1994 for details). For $n \gta 4$, the mode
dissipation time decreases rapidly with increasing mode order (decreasing 
frequency), roughly as $\omega^{-7.5}$. This scaling is explained below 
using a WKB analysis.

The local radiative dissipation rate for the energy of a g-mode is
(Unno et al. 1989) $$\gamma (r) \approx \frac{\Delta
T}{\epsilon(r)T}\frac{\partial \Delta F_R}{\partial r},\eqno(\new)$$
where $\Delta$ denotes a Lagrangian perturbation, $T$ and $F_R$ are
the temperature and radiative flux, respectively, and $\epsilon(r)$ is
the mode's local energy density.
Away from the turning points, this expression reduces to 
$$\gamma (r) \approx \frac{F_{R} k^2_r}{\rho c_s^2 d lnT/dr}
\left(\frac{\partial lnT}{\partial ln \rho} \right)_p
\left(\frac{\partial lnT}{\partial ln \rho} \right)_S,\eqno(\new)$$ 
where $$k_r \approx \frac{1}{\omega_\alpha}\big[N^2 -
\omega^2_\alpha\big]^{1/2}\big[\ell(\ell+1)/r^2 - \omega^2_\alpha/c^2_s 
\big]^{1/2}  \eqno(\new)$$ is the wave's radial wave number and $c_s$ and $N$
are the sound speed and {Brunt-V\"ais\"al\"a~frequency,
respectively.  The global radiative dissipation rate for a $g$ mode
($\Gamma_{\alpha}$) is the integral of $\gamma (r)$, weighted by $\epsilon(r)$,
between the lower and upper turning points:

$$\Gamma_{\alpha} \approx \int^{r_u}_{r_l} dr\, r^2
\gamma(r) \epsilon(r), \eqno(\new)$$
where $r_u$, the outer turning point of the wave, is the radius at
which $\omega^2_\alpha/c^2_s(r_u) = \ell(\ell+1)/r^2_u$.  Energy
conservation implies that $r^2\epsilon(r) v_g$ is independent of $r$
(where $v_g$ is the radial group velocity of the wave) and so the
above equation reduces to

$$ \Gamma_\alpha \approx {F_R \ell \over \bar N \omega_\alpha^2}
\int^{r_u}_{r_l} dr\, {N^3 H\over r^2 \rho c_s^2}\left[
{\ell(\ell+1)\over r^2} - {\omega_\alpha^2\over c_s^2}\right]^{1/2},
\eqno(\new) $$ where $\bar N\equiv \int_{r_l}^{r_u} dr N/r$ 
is the mean value of the {Brunt-V\"ais\"al\"a~frequency.  Most of the
contribution to the wave damping comes from a region near the upper
turning point close to the surface of the star, which moves outward
with decreasing frequency.  Performing this integration for a
polytrope of index 2 (a value appropriate for the outer envelope of
the Yale group's B-star model we are using), we find that
$\Gamma_{\alpha}$ for g-modes scales as $\omega^{-7}$ or $n^7$ (and
also as $\ell^7$).  We note that for a fixed $n$, the value of
$\Gamma_\alpha$ for a quadrupole mode is independent of $\Omega_*$,
$\omega_\alpha$, and $m$ for a uniformly rotating star (which is why
we have used $\ell$ instead of $\lambda$ in the above equations).

\bigskip
\centerline{\bf 2b. Standard dynamical tidal evolution for the PSR J0045-7319
 Binary}
\medskip

The rate of change of the orbital period for the SMC Binary can be
calculated using the following equation
$$ {\dot P_{orb}\over P_{orb}} = -{3\dot E_{orb}\over 2E_{orb}} =
   {3\over 2 E_{orb}}\sum_\alpha \Gamma_\alpha E_\alpha^{(i)} \approx   {3\over 2 E_{orb}}\sum_\alpha {\Gamma_\alpha E_\alpha^{(r)}
    \omega_\alpha^{(i)}\over \omega_\alpha^{(i)} - m\Omega_*}, \eqno(\new)$$
where $\omega_\alpha^{(i)}$ is the mode frequency
in an inertial frame; the frequency in the rotating frame of the star 
is $\omega_\alpha^{(r)} = \omega_\alpha^{(i)} - m\Omega_*$.
The last equality in equation (9) follows from using the relation derived 
earlier between the mode energy in an inertial and rotating reference frame.
The mode energy in the rotating frame of the star,
$E^{(r)}_\alpha$, can be calculated using equation (\energy).
The eigenfrequencies and the overlap of the modes with the tidal
potential ($Q_\alpha$) are calculated using a 10 $M_{\odot}$ B-star
model of 5.3 solar radii (provided by the Yale group). The effect of
the rotation of the B-star on the mode properties is included using the
`traditional approximation' (Bildsten et al. 1996; Chapman \& Lindzen
1970; Unno et al. 1989; see Appendix B for details).  The mass and the
radius of the B-star are observationally estimated to be about $8.8
M_\odot$ and $6 R_\odot$, respectively\footnote{$^3$} {This radius
is slightly smaller than that used by Lai (who took $R_* \approx 6.4
R_\odot$), and so our tidal energies are smaller than his. It can be shown that the energy in the dynamical tide scales as $E^{(i)}_{tide}
\propto Q_\alpha^2\omega_\alpha^2$ (see eq. [3]), where $\alpha$ is the 
mode that carries most of the tidal energy; the frequency of 
this mode in an inertial frame is $\approx 2\Omega_p$. Thus $E^{(i)}_{tide} 
\propto Q_\alpha^2\approx (R_*^5/G)(n_\alpha+1)^{-2\beta}$, where $\beta$ 
is equal to 1.5 (3) for a polytropic star of index 3 (2), and $n_\alpha$ is 
the number of radial nodes for the g-mode with the most energy. 
Since $(n_\alpha+1)
\propto (2\Omega_p)^{-1} \sqrt{GM_*/R_*^3}$, we find that $E^{(i)}_{tide}
\propto R_*^{3\beta+5}/M_*^\beta$; thus for stars of polytropic index 3 (2)
the tidal energy increases with the stellar radius (for fixed stellar mass) as $R_*^{9.5}$ ($R_*^{14}$);
we have verified this dependence using numerical calculations.}
(Bell et al. 1995) and so we multiply the mode frequencies and overlap
integrals calculated
using the Yale model by a factor of 0.78 and 1.36, respectively, to
account for the difference between the stellar model and the
observational parameters; note that $Q_{\alpha}\propto R^{2.5}_*$ and
$\omega_\alpha \propto (M_*/R^3_*)^{1/2}$.  We have neglected the
modification to the equilibrium structure of the star due to rotation in our
calculations, which is likely to be a poor approximation since $\os$
for the SMC B-star is not much smaller than one.

Figure (2) shows $\Delta E_{orb}\equiv P_{orb}\dot E_{orb}$ and
$P_{orb}/\dot P_{orb}$ for the SMC Binary as a function of the Bstar's
rotation rate, $\os$; modes of $m=\pm2$ as well as $m=0$ are included in
these calculations.\footnote{$^4$}{
We note that the results presented in Figures 1 \& 2 were calculated
using slightly more accurate expressions for $E^{(i)}_\alpha$ and 
$E^{(r)}_\alpha$ in which the entire fourier series, not just the resonant
term, is kept (see Appendix A).}  The results presented in this figure, as
well as fig. 1, were obtained by artificially setting $\d =0.2$ in the
calculation of the mode amplitude, thus limiting the strength of
resonances; we will subsequently discuss the probability that closer
resonances significantly enhance the energy in modes and thus decrease
the orbital evolution timescale.  The minima of $E^{(i)}_{tide}$,
$E^{(r)}_{tide}$ and $\Delta E_{orb}$ (and thus the maxima of
$P_{orb}/\dot P_{orb}$) in Figures (1) and (2) correspond to the
`off-resonance' dynamical tide.  They are equivalent to the
calculation of the work done on the star in the absence of any
oscillations, i.e., the result obtained by Press \& Teukolsky 1975
(see also Lai 1996).

For $\os < 0$ the tidal frequency is larger than for prograde
rotation, and so the frequencies of the modes with the most energy are
higher and their $n$ value is lower. Thus the dissipation time of the
dynamical tide is longer for retrograde rotation than for prograde
rotation.  For example, for $\os \approx -0.3$, the modes with the
most energy are the $g_1$-$g_3$ modes, while for $\os \approx 0$, the
modes with the most energy are the $g_4$-$g_8$ modes.  The dissipation
times are longer in the former case by a factor of about 100 (see
Table 2).  This is why, although the energy in the dynamical tide is
about 100 times larger for retrograde rotation than for no rotation (fig. 1),
the energy dissipated, and thus the orbital evolution time, is
comparable in the two cases.  We note that $\omp/2\pi \sim 3.6 \mu$Hz,
which accounts for the minimum in $\Delta E_{orb}$ and the maximum in
$P_{orb}/\dot P_{orb}$ at $\os\approx 0.2$ (see fig. 2), since this
is where $\Omega_{tide} \approx 0$. We would also like to point out that
the energy in the dynamical tide for $\os \approx 0$ is a few times
10$^{40}$ erg (fig. 1). If this energy is dissipated in an
orbital period, the resulting orbital period evolution time 
is about 10$^6$ years, which is close to the observed value.

The observations of spin-orbit coupling in the timing data of PSR
J0045-7319 provides evidence that the magnitude of the
B-star's rotation rate is probably super-synchronous, that is, $|\Omega_*|\gta
\Omega_p$ (Lai et al. 1995; Kaspi et al. 1996).  Prograde rotation at
$\Omega_* \gta \Omega_p$ would lead to a transfer of energy from the
spin of the B-star to the orbit, causing the orbital period to
increase with time (see fig. 2 and the last part of the discussion following eq.
[\energy]), which is inconsistent with the observations; prograde
super-synchronous rotation of the B-star is thus ruled out. For $\Omega_*\sim\Omega_p$ 
(which is consistent with the apsidal motion of the binary system)
the energy in the dynamical tide is too small by about two orders
of magnitude to explain the observed orbital evolution time,
even if we assume that all of the tidal energy is dissipated in one orbit.
These considerations force us to conclude that the rotation of the B-star
must be retrograde. This is discussed further in \S3.

Our calculations indicate that the orbital evolution time due to the
radiative dissipation of the `off-resonance' dynamical tide is at
least three orders of magnitude too small to account for the observed
$P_{orb}/\dot P_{orb}$ of the PSR J0045-7319 binary, regardless of the
B-star's rotation rate.  Is it possible, however, that the observed
orbital decay rate of the PSR J0045-7319 Binary is because a low to
moderate order $g$ mode is highly resonant with the orbit ($\dr\approx 
10^{-2}$), thus increasing the energy in the dynamical tide by about three
orders of magnitude and decreasing the orbital evolution timescale to
about $5 \times 10^5$ years?  We think that this is highly unlikely.
If the `off-resonance' value of the energy in the dynamical tide is
$E_{or}$, then the probability that we are observing the system at a
time when the energy in the dynamical tide is $E$ is $P(E) \approx
N(E_{or}/E)^{3/2}$, where $N$ is the number of low to moderate order
$g$ modes which can be resonant with the orbit.  This follows from
comparing the time it takes the resonant mode to move off resonance
with the time it takes the orbit to evolve between resonances.  From
Figures (1) and (2), $E/E_{or} \approx 10^3$ in order to explain the
observations and thus the probability that the orbital decay of the
PSR J0045-7319 Binary is due to a resonant $g$ mode is $\sim 0.01 \%$.
This probability is nearly the same for $\os = - 0.3, 0$.  We thus
conclude that it is highly unlikely that the orbital decay of the PSR
J0045-7319 Binary is due to the radiative dissipation of the dynamical
tide in a uniformly rotating star.  We now consider several alternate
dissipation mechanisms, including the effect of differential rotation
on the damping of gravity waves.

\bigskip
\centerline{\bf \S3. Alternative damping mechanisms for tidal waves}
\medskip

In the previous section we have argued that the linear radiative
dissipation of the dynamical tide in a uniformly rotating B-star is
incapable of explaining the observed orbital decay of the SMC Binary.
The calculations of the energy in the dynamical tide are, in our opinion,
relatively secure, but the understanding of dissipation mechanisms, 
which clearly affects our estimate of the orbital evolution time, requires 
closer scrutiny. Accordingly, we describe two dissipation mechanisms in 
this section. One of them, described below, is found to be extremely
efficient in dissipating the energy of the dynamical tide when the
star has some differential rotation. Another mechanism, nonlinear
parametric coupling of the equilibrium tide to low frequency g-modes,
is more likely to operate when the star is rigidly rotating. This is 
discussed in \S3b.

\bigskip
\centerline{\bf 3a. The effect of differential rotation on the radiative 
damping of gravity waves}
\medskip

The unusual (non-periodic) time residuals for the pulse arrival times
of PSR J0045-7319 provides strong evidence for spin-orbit coupling in
the PSR J0045-7319 Binary and suggests that the interior of the B-star
is rapidly rotating with its spin axis inclined with
respect to the normal to the orbital plane (Kaspi et al. 1996; Lai et
al.  1995).  We can readily estimate the relative timescales for spin
pseudo-synchronization and orbital circularization for the SMC system
to see if the B-star should, from a theoretical point of view, be
rotating pseudo-synchronously (i.e., with $\Omega_* \approx
\omp$).\footnote{$^5$}{The relative timescale calculation is
significantly more secure than a calculation of the absolute timescale
for pseudo-synchronization since it depends only on the ratio of the
energy to the angular momentum in the tide.} 

The rate of change of the stellar rotation due to the transfer of orbital
angular momentum to the star is $|\dot\Omega_*|/\Omega_* \approx
(L_{orb}/L_*) \ (|\dot L_{orb}|/L_{orb})$, where $L_{orb}/L_*
\approx 4 \hat \Omega^{-1}_*$. Using $\dot L_{orb} = \dot E_{orb}/\Omega_p$
for the dynamical tide (see $\S2$ and appendix A), as well as for the 
equilibrium tide\footnote{$^6$} {The equilibrium tide in the weak friction
limit is discussed in detail in the seminal paper of Hut (1981), who also
finds $\dot E_{orb} \approx \Omega_p\dot L_{orb}$.}, we find that
$|\dot \Omega_*|/\Omega_* \approx 0.2~\hat\Omega^{-1}_* (|\dot
E_{orb}|/E_{orb})$. Thus so long as the B-star is rotating near break
up ($|\os|\gta 0.2$), {\it both} the equilibrium and the dynamical
tides lead to timescales for spin pseudo-synchronization which are
comparable to the orbital circularization time.  We therefore do not
expect that the rotation rate in the interior of the B-star
has changed appreciably, which is consistent with the observations.

We do expect, however, that the dynamical tide should have forced
the surface layers of the B-star into pseudo-synchronous rotation.
The physical reason for this
follows from the seminal work of Goldreich and Nicholson (1989), who
showed that tidal waves deposit their angular
momentum at the place in the star where they are dissipated. Since the
dissipation rate is largest near the surface of the star (see \S2b)
and the moment of inertia of the surface region is a small fraction of
the star's total moment of inertia, the surface of the star tends to
be pseudo-synchronized very rapidly.  For the SMC binary system with a
rapidly retrograde rotating B-star ($\os \approx -0.3$), the modes
with the most energy have frequencies $\approx 17 \mu$Hz in the
rotating reference frame.  Figure (3) shows $\gamma(r)$, the local
radiative dissipation rate, for such a mode.  The dissipation is
concentrated in a layer $\approx 0.05R_*$ thick near the outer turning
point, which occurs at $r \approx 0.85 R_*$.  The moment of inertia of
this layer is $\approx 10^3$ times smaller than that of the entire
star and so we expect the rotation of the B-star at $r \approx 0.85
R_*$ to be pseudo-synchronous. The dominant tidal waves are, however,
evanescent at $r\gta 0.85 R_*$, and
lower frequency waves with outer turning points near
the surface of the star ($\omega_\alpha/2\pi \lta 5 \mu$Hz) carry
insufficient angular momentum to bring the surface ($r\gta 0.85R_*$)
into pseudo-synchronous rotation. The large positive entropy gradient
in the radiative exterior of the B-star (the Brunt-V\"ais\"al\"a~frequency is 
$\approx 70 \mu$Hz) has a strong stabilizing effect on 
shear instabilities which could otherwise retard the surface rotation
and reduce the differential rotation 
of the region at $r\approx 0.85 R_*$. Therefore, provided that magnetic
stresses do not efficiently transport angular momentum between the 
stellar surface and the pseudo-synchronous layer, the rotation rate of 
the B-star at $r \gta 0.85 R_*$ is expected to be approximately the same as 
it was at the initial time.
The optical linewidth measurement of Bell et al. (1995) gives a 
surface rotation frequency of $\approx 4/\sin(i_{ns}) \ \mu$Hz, where 
$i_{ns}$ is the angle between the spin axis and the line of sight.  
If correct, this suggests that the observed surface of the star is perhaps 
not pseudo-synchronized.
In what follows, we shall assume that there is a region below the 
surface of the star that rotates close to the pseudo-synchronous value and
describe how this dramatically enhances the dissipation of the dynamical tide.

We have calculated g-mode eigenfunctions in rigidly and differentially
rotating stars and an example of the adiabatic energy flux for a
g-mode is shown in Figure (4).  The differential rotation was chosen
to conform to that expected from the radiative dissipation of the
dynamical tide in the B-star of the PSR J0045-7319 Binary, i.e., the thickness
of the differentially rotating layer was taken to be $\approx 0.1 R_*$ 
centered at $r \approx 0.8 R_*$.  We note that the energy flux is nearly
constant across the differentially rotating layer, indicating that
there is very little reflection of the wave at this layer; furthermore,
the wavelength of the wave has decreased dramatically in the
differentially rotating layer, which results in a strongly enhanced
dissipation rate.

The frequency of tidally excited gravity waves, as seen by an inertial
observer, is $\approx2\Omega_p$ (see \S2a). If the component of the 
rotation rate of
the star normal to the orbital plane a distance $r$ from the center is 
$\Omega_*(r)$, then the
frequency of the gravity wave in the local rest frame of the fluid is
$\omega_r \approx2|\Omega_*(r) - \Omega_p|$ and its wavenumber is
$\approx6^{1/2}N(r)/(r\omega_r)$. Thus, as the wave approaches the
pseudo-synchronously rotating surface of the star its wavelength goes
to zero and the energy it carries is entirely
dissipated.\footnote{$^7$}{As the wave frequency goes to zero, the
effect of the Coriolis force on the wavefunction becomes increasingly
important and it might appear that this will adversely affect the
dissipation of the wave. Using the `traditional approximation'
(Appendix B), however, it can be shown that, for a fixed wave
frequency and a surface rotation rate which is slightly
sub-synchronous, including the Coriolis force tends to increase the
wavelength of the wave by only $\approx$ 20\%, and thus the dissipation of the wave
remains essentially the same.} Even if the rotation of the B-Star
in the SMC Binary at $r \approx 0.85 R_*$ is not fully pseudo-synchronized,
the dynamical tide will be completely absorbed as it approaches the
surface provided that its local radiative dissipation rate becomes
comparable to the wave frequency.  We find that this occurs so long as
the frequency of the gravity wave in the local rest frame of the star
near $r \approx 0.85 R_*$ is less than about $2 \mu$Hz, which requires
that the stellar rotation frequency at $r \approx 0.85 R_*$ be within about
$1 \mu$Hz (or 30\%) of its pseudo-synchronous value. This
requirement, which is not particularly stringent, is weakened further
if waves can travel closer to the surface of the star, such as when
there is redistribution of angular momentum due to internal stresses
in the outer envelope of the star.

Let us assume that the energy input into the waves, inspite of the
enhanced dissipation, is approximately equal to the energy in tidal 
oscillations calculated in \S2a in the absence of resonances (the error due to
this assumption is discussed below). From Figure 1 we see that this energy 
input is $\approx 2
\times 10^{41}$ ergs per orbit when the star is rotating with $\hat
\Omega_*=-0.3$ ($\Omega_*/2\pi \approx -6 \mu$Hz).  
Using $P_{orb}/\dot P_{orb} = -2 E_{orb}/3\dot E_{orb}$ with an energy
loss per orbit of $\Delta E_{orb} \approx E^{(i)}_{tide}
\approx 2 \times 10^{41}$ ergs, we find an orbital period evolution time for 
the SMC binary of $\approx$ $10^5$ years, which is a factor of about 5 smaller 
than the observed timescale.

Thusfar we have ignored the modification to
the g-mode eigenfunction in the differentially rotating star.  As
Figure (4) shows, in the outer part of the star where the stellar
rotation rate approaches its pseudo-synchronous value, the wavelength
of the gravity wave becomes very small.  Thus, the coupling of gravity
waves to the tidal forcing function is reduced, i.e., there is very
little contribution to the overlap integral from the outer part of the
star ($r\gta 0.8 R_*$).  The exact amount by which the overlap
integral is reduced in the differentially rotating star depends on the
details of the differential rotation and the modes which are excited,
and is somewhat uncertain.\footnote{$^8$}{When the tidal frequency
is much less than quadrupole f-mode frequency, almost all of the
forcing by the tidal potential occurs near the interface of the
convective core and the radiative exterior where the wavelength of the
wave is the longest.  Thus, differential rotation in the outer part of
the star would not substantially modify the coupling of the wave to
the tidal force. This limit is, however, not applicable to the very low order
modes excited in the rapidly retrograde rotating B-star.} Our estimates
suggest that for the very low order modes excited in the rapidly
retrograde rotating B-star, $Q_\alpha$ decreases by a factor $\sim
2$ due to the differential rotation and so the energy in the dynamical
tide decreases by about a factor $\sim 4$. This increases our
estimate of the orbital period evolution timescale of the PSR J0045-7319
Binary to $\approx 4$x10$^5$ years, which is consistent with the
observations. In fact the theoretically calculated orbital evolution
times are consistent with the observations so long as $\-0.3\lta\hat\Omega_*\lta 0$.

We note that prograde rotation of the B-star at $\hat \Omega_* \approx
0.4$ gives an orbital evolution time of comparable magnitude, though
with the opposite sign. Thus our reason for suggesting retrograde
rotation is not to explain the short timescale for orbital evolution
(which prograde rotation can also do), but rather to insure that the
tidal interactions result in a decrease in the orbital period with
time, as is found by the observations. For $\hat\Omega_*\sim 0.1$ (which
is roughly the observational lower limit obtained from the apsidal motion
of the system, cf. Kaspi et al. 1996)
the energy in the dynamical tide is less than a few times 10$^{39}$ erg; 
the resulting orbital evolution time, assuming that all of this energy is 
dissipated in one orbit, is greater than about $3 \times 10^7$ years,
which is a factor of $\approx$ 50 larger than the observed value. 
For $\hat\Omega_*\approx 0$, on the other hand, the energy in the tide is
about 2x10$^{40}$ erg and thus the orbital period is expected to 
decrease on a time scale of $\sim 10^6$ years. This small rotation rate
is, however, inconsistent with the observed apsidal motion of the system
(Kaspi et al. 1996), which is why we are forced to consider retrograde 
rotation. 

It is straightforward to show that the rate of change
of the orbital eccentricity is given by $\dot e\approx 
\bigl[(1-e^2)/3e\bigr](\dot P_{orb}/P_{orb})$. Thus we expect 
$\dot e^{-1}\approx$ 3x10$^6$ years, which is a factor of about 20 larger
than the current lower limit of Kaspi et al. (1996).

If magnetic stresses and/or instabilities are very efficient in
redistributing angular momentum in the B-star so that, inspite of the
tidally excited waves depositing angular momentum at $r\sim 0.85 R_*$,
the star continues to rotate almost rigidly, the dissipation of gravity
waves will not be enhanced as described above. However, tides may still be
efficiently dissipated by the nonlinear process discussed by Kumar and
Goodman (1996). This is considered in some detail below.

\bigskip
\centerline{\bf 3b. Parametric coupling of the dynamical and equilibrium 
tides with g-modes}
\medskip

When a primary mode, denoted by $\alpha$ and taken to be either the
dynamical or the equilibrium tide, drives a mode of half its frequency
(referred to as the daughter mode and denoted by $\beta$) by nonlinear
mode coupling, the process is called parametric instability. This
process was proposed by Kumar \& Goodman (1996) for damping the
dynamical tide in late type stars, and was found to be much more
efficient than conventional turbulent and/or radiative dissipation 
for close binary systems. We consider this process for the B-star of the 
SMC binary.

The growth rate of the energy of the daughter mode, subject to parametric
driving, is given by (Kumar \& Goodman, 1996) 
$$\eta =  \left[ \sqrt{18 E\kap^2
\omega^2_{\beta} - (\Delta\omega)^2} - {\Gamma_{\beta}\over 2}
\right], \eqno(\new)$$ 
where $E$ is the energy in the primary mode, $\Delta
\omega \equiv \omega_\alpha - 2\omega_\beta$, $\Gamma_\beta$ is the linear 
energy dissipation rate of the daughter mode, and $\kap$ is the
nonlinear 3-mode coupling coefficient (the calculation of which is
somewhat involved and is described in Kumar \& Goodman).  Thus,
parametric instability sets in only if there are modes in the star which
can simultaneously satisfy: $$|\Delta \omega| <
3\sqrt{2}E^{1/2}\kap\omega_\beta \eqno(\new)$$ and $$\Gamma_\beta <
6\sqrt{2}E^{1/2}\kap\omega_\beta. \eqno(\new)$$ These two conditions
compete in the sense that higher $\ell$ g-modes have smaller frequency 
spacings and are thus more likely to
satisfy equation (\ref2), but they have larger dissipation rates and are
thus less likely to satisfy equation (\last).

The damping rate, $\Gamma_\beta$, for g-modes of early type stars
increases as $\ell^7$ (see \S2a), and so only 
low $\ell$ daughter modes can satisfy equation (\last). For retrograde
rotation of $\hat \Omega_* = -0.3$, the energy in the primary modes
($g_2-g_4$) is $E \approx 2 \times 10^{41}$ ergs (see Fig. 1), and the
3-mode coupling coefficient of these modes with daughters of half
their frequency is $\kap \approx 10^{-25}$ erg$^{-1/2}$. Thus, we must
find a daughter mode with $|\Delta \omega|/\omega_\beta \approx 10^{-4}$ 
and $\Gamma_\beta \lta 3$x$10^{-8}$ s$^{-1}$ in order for the
dynamical tide to be  parametrically unstable. The small damping 
time required for the
daughter modes implies that it must be of degree less than $\sim 3$.
However, since the g-mode frequencies in the B-star of the
SMC binary are $\sim (n+1)^{-1} \sqrt{\ell(\ell+1)} \ 27 \
\mu$Hz, it is very unlikely (the probability is less than $\approx$ 0.1\%) 
that there
is such a low degree daughter mode with a frequency within the required
tolerance that can couple to the primary mode.  We
thus conclude that the dynamical tide in the PSR J0045-7319 Binary is
stable against parametric instability. We show below, however, that the 
equilibrium tide is subject to the parametric
instability.

The equilibrium tide represents the hydrostatic response of the
primary to the perturbing gravitational force of the secondary.  The
structure of the equilibrium tidal perturbation is similar to that of
the $f$ mode.  The frequency of the equilibrium tide, $\Omega_{tide}
= 2|\omp - \Omega_*|$, is, however, much smaller than that of the $f$
mode. Thus the nonlinear coupling of the equilibrium tide to its
daughter modes is analogous to the coupling of the $f$ mode to very
low frequency ($\omega_\beta \approx \Omega_{tide} \ll \omega_f$)
daughter modes, rather than to daughter's of half its frequency, as
was the case for the dynamical tide.  For $\omega_\beta \ll \omega_f$,
it can be shown that the coupling coefficient is proportional to the
square of the daughter's normalized transverse displacement
eigenfunction, which is proportional to $\omega^{-2}_{\beta}$.  Thus,
the coupling coefficient for the equilibrium tide is significantly
larger than for the dynamical tide.  For $\ell \approx 1$ daughter
modes, $\kap \approx 4 \times 10^{-22} \nu^{-2}_\beta$~erg$^{-1/2}$,
where $\nu_\beta=\omega_\beta/2\pi$ is the daughter mode frequency in
micro-Hz.\footnote{$^9$}{ For $\nu_\beta = 30 \mu$Hz (half the $f$ mode
frequency), this gives $\kap \approx 10^{-25}$erg$^{-1/2}$, which agrees
well with the nonlinear coupling of the low order dynamical tide given 
above, as is expected since the equilibrium tidal perturbation is 
similar to that of the $f$ mode.}

The energy in the equilibrium tide near periastron is $E_{eq} \approx
k (R_*/R_{peri})^6 GM^2/R_* \sim 10^{42}$ ergs, where $R_{peri}
\approx 4 R_*$ is the separation between the two stars at
periastron and $k \sim 0.008$ is the apsidal motion constant of the
B-star.  The energy in the equilibrium tide falls off rapidly from
periastron, so parametric instability can only set in and daughter
modes can only grow near periastron.  We thus define an orbit averaged
growth rate for a daughter mode as $\bar \eta \equiv \eta \om/\omp$,
where $\eta$ is calculated using the energy in the equilibrium tide at
periastron. If $\bar \eta > \Gamma_\beta$, the daughter mode has net
growth over an orbit.

For rapid prograde rotation of the B-star ($\Omega_*>\Omega_p$) the
dissipation of the equilibrium tide causes the orbital period to
increase with time, as was the case with the dynamical tide discussed
in \S2 (see appendix A and Hut 1981 for a detailed discussion).  Thus,
only nonlinear dissipation of the equilibrium tide in a rapidly
retrograde rotating B-star can potentially explain the observed orbital
period decrease of the SMC binary.

For the SMC binary system, using the values given above for $E_{eq}$
and $\kap$ (and taking $\nu_\beta \approx 8 \mu$Hz, appropriate for
retrograde rotation at $\os \approx -0.3$) we estimate that $\eta
\approx 10^{-6}$ s$^{-1}$ and $\bar\eta\approx 10^{-7}$ s$^{-1}$.
Thus we need to find daughter modes within about 0.2 $\mu$Hz of 8
$\mu$Hz which is possible for $\ell_\beta \ge 2$ when we consider the
lifting of m-degeneracy by stellar rotation. We also find that $\bar
\eta > \Gamma_\beta$ so long as $\ell_\beta \lta 4$ (using 
$\Gamma_\beta\propto \ell_\beta^7$ and the damping of quadrupole modes
given in Table 2).  Thus, we conclude that the equilibrium tide in the
SMC Binary is subject to parametric instability which, as we see
below, results in a rather efficient dissipation of its energy.

The mean growth time of these daughter modes is $\bar\eta^{^{-1}}\approx$
100 days. The resulting nonlinear dissipation time for the equilibrium
tide ($t_{nl}$) depends on the energy in the daughter modes, 
which is expected to be comparable to the energy in the
equilibrium tide since the daughter modes are not likely to lose energy
to granddaughter modes via parametric instability. Thus, $t_{nl} \approx 
\eta^{-1}$.  The orbital evolution timescale which results from the 
nonlinear dissipation of the equilibrium tide is $P_{orb}/\dot P_{orb}
\approx (E_{orb}/E_{eq}) \ (\omp/\om) \ t_{nl} \sim 
\bar\eta^{^{-1}} (E_{orb}/E_{eq})\sim 10^5$ years, in reasonable agreement 
with the observations.

We feel, however, that the case for nonlinear damping in the SMC
Binary, while encouraging, is not entirely robust. This is because the
nonlinear growth rate of the daughter modes is comparable to their
linear radiative damping rates and because there are not many daughter
modes available for coupling with the equilibrium tide. Thus a change
by a factor of a few in either $\kap$ or $\Gamma_\beta$, which is
within the uncertainty of the stellar model, could modify our conclusion.

\bigskip
\centerline{\bf 4. Discussion}
\medskip

The PSR J0045-7319 Binary in the SMC consists of a radio pulsar in a
51 day, highly eccentric (e=0.808), orbit with a B-star companion of
mass $\approx 8.8 M_\odot$. The orbital period of this system is
observed to be decreasing on timescale of $\approx 5$x$10^5$ years
(Kaspi et al.  1996), which is several orders of magnitude faster
than that expected from the standard theory of the dynamical tide.  Timing
observations of PSR J0045-7319, as well as optical linewidth measurements, 
show that the B-star is rapidly rotating with a rotation speed that
is perhaps greater than the orbital angular speed of the star at
periastron (Lai et al.  1995, Kaspi et al. 1996, Bell et al. 1995).

We find that for rapid prograde rotation of the B-star (rotation
frequency greater than the orbital angular speed of the star at
periastron) the dissipation of both the equilibrium and dynamical
tides lead to an increase in the orbital period of the binary system.
This is because the tidal torque is slowing down the spin of the star
and the resultant loss of rotational kinetic energy exceeds the energy
deposited in the tide.  Thus, we infer that the B-star must have
retrograde rotation if tidal effects are to account for the observed
period decrease of the PSR J0045-7319 Binary.  Lai (1996) has
suggested that rapid retrograde rotation of the B-star can explain the
orbital evolution. We certainly agree with this conclusion. However,
our reason for requiring retrograde rotation is entirely different
from his. This point is clarified below.

We find that for $\os\equiv \Omega_*(R^3_*/GM_*)^{1/2}=-0.3$ (0.0) the
energy in the dynamical tide, as seen from an inertial frame, is about 
2x10$^{41}$ (2x10$^{40}$) erg and the number of radial nodes for the most 
excited g-mode is 3 (6).\footnote{$^{10}$}{Our energies are smaller than 
those of Lai in part because we use different reference frames and stellar 
models, and also because we use a somewhat smaller stellar radius 
(see the footnote in \S2b).} The much larger 
energy in the dynamical tide for rapid retrograde rotation was 
correctly pointed out by Lai. 
However, he took the damping time of g-modes to be independent of frequency 
(with a value of about 14 years) and concluded that the increase in the 
tidal energy for rapid retrograde rotation could account for the
observed rapid evolution of the PSR J0045-7319 Binary.  We have shown,
however, that for a rigidly rotating star the mode dissipation time 
increases rapidly with frequency (roughly as $\omega^{7}$) and so the 
high frequency modes excited in the retrograde rotating B-star are much 
less efficiently damped than the lower frequency modes excited in a non-rotating
B-star.  The damping time of the dynamical tide, in a rigidly rotating
star, for $\os=-0.3$ (0.0) is given in Table 2 to be about $1.1$x10$^3 (50)$ 
years (the dissipation was calculated by solving the fully
nonadiabatic oscillation equations and the results are consistent
with, for example, those of Saio \& Cox 1980).  Thus the orbital
evolution time for rapid retrograde rotation of the B-star, inspite of
the enhanced energy in the tide, is the same as for no rotation and is
$\approx$ 10$^9$ years, or a factor of 10$^3$ longer than the observed
value.  We thus emphasize that retrograde rotation of the B-star is
required, not because the standard theory for the dissipation of dynamical 
tide in a retrograde rotating star can explain the observed  
 period decrease of the
PSR J0045-7319 Binary (as Lai suggested), but rather because rapid
prograde rotation of the B-star would lead to orbital period growth,
which is inconsistent with the observations. In fact, with the enhanced
dissipation mechanisms discussed below,  the orbital evolution timescale
for $\hat\Omega_* \approx +0.4$ is also consistent with the observations, 
but of course in this case the orbital period increases with time.

The energy in the dynamical tide for $\os\approx -0.3$ is about 10$^{41}$ erg
(fig 1b). If this energy is dissipated in about an orbital period then
the resulting theoretical time scale for the evolution of the orbital period
is similar to the observed orbital decay time for the SMC Binary.
We have considered two different dissipation mechanisms that 
satisfy this property. One of these
requires significant differential rotation of the B-star, which is
expected when a star is not pseudo-synchronized. This is because the
gravity waves deposit their angular momentum in the outer layers of
the B-star where the dissipation occurs and where the moment of
inertia is much smaller than in the interior (Goldreich \& Nicholson
1989).  For $\os=-0.3$, the dissipation of the
dynamical tide in the B-star is concentrated at $r \approx 0.85 R_*$ and so, 
provided that magnetic torques and/or instabilities do not force this
region into solid body rotation with rest of the star, this region
should be rotating pseudo-synchronously (i.e., with a rotation
frequency close to the orbital angular frequency of the star at
periastron, $\omp/2\pi \approx 3.6 \mu$Hz).

The frequency of the dynamical tide in an inertial reference frame is
$\approx 2\Omega_p$ and so the frequency in the local rest frame of
the star is $\approx 2[\Omega_p - \Omega_*(r)]$, where $\Omega_*(r)$
is the component of the stellar rotation rate at radius $r$ normal to the 
orbital plane. Thus, the frequency and
wavelength of the dynamical tide in the local rest frame of the star
decrease as the wave approaches the psuedo-synchronously rotating layer.
The decreasing wavelength results in the entire gravity wave energy flux 
being dissipated by radiative diffusion near
the surface of the star, provided that the rotation frequency of the star at 
$r \approx 0.85 R_*$ is within about 30\% of the pseudo-synchronous value.
This enhanced dissipation readily accounts for the observed orbital period 
decrease of the PSR J0045-7319 Binary so long as $-0.3\lta\hat\Omega_*\lta 0$.
The resulting timescale for the 
evolution of the orbital eccentricity (1/$\dot e$) of the PSR J0045-7319 
Binary is $\approx$ 3x10$^6$ years, which is a factor of 
about 20 larger than the lower limit of Kaspi et al. (1996).

The observed orbital period evolution can also be understood provided 
that the interior of the B-star is rotating very slowly
(with the exterior close to the surface rotating pseudo-synchronously).\footnote{$^{11}$}{Figure
1b shows that the tidal energy for a slowly rotating B-star,
$\os\approx 0$, is about 10$^{40}$ ergs, which is a factor of $\sim$5 smaller 
than what is needed to explain the orbital evolution. The tidal energy,
however, depends on the radius of the B-star, which is not precisely known;
thus, if the radius of the star is some what larger than the value we have 
chosen, the tidal
energy will be larger than the value shown in figure 1.}
Slow rotation of the B-star's interior ($|\hat\Omega_*| \lta 0.1$) is, 
however, 
ruled out by the apsidal motion observations of Kaspi et al. (1996). 
Thus retrograde rotation of the B-star with $\hat\Omega_*$ between 
$\approx$ -0.3 and $\approx$ -0.1 appears to be the only solution that fits
all of the current observations of this binary system. 

In the likely scenario that the binary system was synchronized and circularized
prior to the supernova that created the pulsar, the  
current retrograde rotation of the B-star implies that the
supernova had a dipole asymmetry which imparted a net angular
momentum kick to the system. This point was also made by Kaspi et al.
on the basis of the misalignment of the spin and angular momentum axes.

We note that in order for this mechanism to work the B-star must
have a large gradient of differential rotation in the outer envelope,
with the specific angular momentum decreasing outward in the star.
This can lead to thermal instabilities such as the
Goldreich-Schubert-Fricke instability, and it is unclear to us whether
angular momentum will be mixed efficiently as a result (the star is,
however, dynamically stable because of the large positive entropy
gradient in the radiative exterior).  This is an important issue which
we have not addressed and which needs to be investigated by a careful
nonlinear calculation.

Another dissipation mechanism we have considered is nonlinear
parametric instability of the equilibrium tide, which can operate when
the differential rotation in the star is small. In this case
energy is transferred, via resonant 3-mode coupling, from the large
scale perturbation associated with the equilibrium tide to low degree
g-modes of half the tidal frequency; the timescale for this process is
found to be $\sim 10^2$ days for the B-star of the SMC binary and the 
resulting orbital evolution timescale is a few
hundred thousand years. The margin of instability in our calculation
is, however, small and so we are not confident that this
dissipation process is operating in the B-star of the SMC 
binary system.

\medskip
\ni{\bf Acknowledgment:} We thank Dong Lai for sending us his paper
prior to publication which we found helpful to our thinking about this 
problem. We are very grateful to Jeremy Goodman for numerous discussions, for
sharing with us his insight on nonlinear damping of the equilibrium tide that 
led to \S3b, and for suggesting a number of improvements to 
our presentation. Obviously it is not his fault if, inspite of this,
there are mistakes in the paper. We also thank Vicky Kaspi for comments.

\vfill\eject
\bigskip
\centerline{\bf Appendix A}
\medskip
In this appendix we calculate the relationship between the angular
momentum and energy deposited by the tidal force for the dynamical and
equilibrium tides.  
The quadrupole tidal gravitational potential, as seen by an observer
corotating with the primary star, is given by $$U = - \beta r^2 \sum_m
Y^*_{2 m}(\theta, \phi) \tilde{f}^*_{2 m},
\eqno(A1)$$ where $$\beta = \frac{4 \pi}{5} \Omega^2_o \frac{M}{M+M_*} 
\eqno(A2)$$ and $$\tilde f_{2m} = \frac{Y^*_{2m}(\pi/2, \phi_{orb}(t) -
\Omega_*t)}{(R(t)/a)^3}. \eqno(A3)$$  The torque exerted on the primary
 star by the tidal force is only in the $\hat z$ direction (taking,
for simplicity, the rotation axis to be perpendicular to the orbital
plane) and is given by $$\tau_z = -\int d^3x ({\bf r} \times \nabla
U)_z \delta \rho,
\eqno(A4)$$ where $\delta \rho$, the Eulerian perturbation to the density
of the star, can be expanded in normal modes as $\delta \rho =
\sum_\alpha A_\alpha \delta \rho_\alpha Y_{\ell m}(\theta, \phi)$.  
Substituting the tidal gravitational potential (eq. [A1]) into
equation (A4), we get $$\tau_z = - i\beta\sum_\alpha m A_\alpha
Q_\alpha \tilde{f}^*_{2 m}. \eqno(A5)$$ 
The non-dimensional function $\tilde{f}_{2 m}$ can be expanded in a 
fourier series as (Kumar et al. 1995; Quataert et al. 1996) 
$$\tilde{f}_{2 m} = \exp(i m \Omega_* t) 
\left[ \sum_{k=1}^{\infty}C^{2 m}_k 
\sin(k\om t) + \sum_{k=0}^{\infty} D^{2 m}_k \cos(k\om t) \right ], 
\eqno(A6)$$  where Im$(D^{2 0}) = C^{2 0} = 0$, Im$(D^{2 \pm 2}_k) = 
$Re$(C^{2 \pm 2}_k) = 0$, and, for $k \gta$ a few, Re$(D^{2 \pm 2}_k) 
\approx - {\rm sign}(m) $Im$(C^{2 \pm 2}_k)$.  
Using equation (A6) one can expand the mode amplitude 
in a fourier series and obtain (c.f. Kumar et al. 1995) 
$$\eqalign{A_\alpha = & \frac{\beta Q_\alpha \omega^2_\alpha}{2 \Omega^2_0} 
\exp(i m \Omega_* t)\sum_k\biggl[ (D^{2 m}_k + iC^{2 m}_k) 
\frac{\exp[-ik\Omega_0 t - 
i\phi_k(m)]}{\eta_k(m)} \cr & \quad \quad + (D^{2 m}_k - iC^{2 m}_k) 
\frac{\exp[ik\Omega_0 t + 
i\phi_k(-m)]}{\eta_k(-m)}\biggr]} \eqno(A7)$$ where $\eta^2_k(m) =
[\r^2 - (k - ms)^2]^2 + \d^2(k - ms)^2$ and the other symbols are the
same as defined in equation (2) of the main text.  Note that
$A_\alpha$ given above is as seen by an observer corotating with the
star. In what follows, we shall assume that Re$(D^{2 \pm 2}_k) = -
{\rm sign}(m) $Im$(D^{2 \pm 2}_k)$ for all $k$.  We find that this
yields values for the energy and angular momentum deposited in the
star which are accurate to within at least 5 \%.  Substituting the
fourier series expansions of the mode amplitude and the tidal forcing
function into the expression for the tidal torque [A5], we find that
the angular momentum deposited in the star, in one orbit, is given by
$$\Delta L \equiv \int^{2\pi/\Omega_o}_0 dt \tau_z = -\frac{8 \pi
\beta^2}{\Omega^2_o} \sum_{n} Q^2_{n 2 2} \omega^2_{n 2 2} d_{n 2 2} \sum_k 
\frac{(D^{2 2}_k)^2}{[\eta_k(2)]^2} (2 \Omega_* - k),
\eqno(A8)$$ where we have summed over $m = \pm 2$ in arriving at the above
expression.  We note that $m/\omega_\alpha$ has the same sign for $m =
\pm 2$ modes (see eq. [A7] or eq. [2] of the main text) and so the modification
to $\omega_\alpha$ and $Q_\alpha$ by the Coriolis force is the
same for these modes (see Appendix B).

The energy input rate into the star is given by $$\dot E = -\int d^3x
\rho \nabla U \cdot {\bf v}, \eqno(A9)$$ where ${\bf v}$, the fluid velocity
in the star, depends on the reference frame.  In an inertial
reference frame, ${\bf v_i} = {\bf v_r} + {\bf \Omega_*} \times {\bf
r}$, where ${\bf v_r} =
\sum_\alpha dA_\alpha/dt \  \bxi_\alpha \  Y_{\ell m}(\theta, \phi)$
is the fluid velocity as seen by an observer corotating with the star
(and is due to the tidally excited oscillations) and ${\bf
\Omega_*} \times {\bf r}$ is the fluid velocity due to the rotation of 
the star (as seen by an inertial observer).  It is straightforward to
show from equations (A4) and (A9) that the energy input as seen by an
inertial observer, $\dot E^{(i)}_{tide}$, is related to the energy input
as seen by an observer corotating with the star ($\dot E^{(r)}_{tide}$) by
$\dot E^{(i)}_{tide} =
\dot E^{(r)}_{tide} + \Omega_* \tau_z$.  The change in the orbital energy per 
orbit includes both the energy input in the modes, $\dot E^{(r)}_{tide}$,
and the changing rotational energy of the star ($\Omega_* \tau_z$) and
is thus given by $$\Delta E_{orb} = -\int^{2\pi/\Omega_o}_0 dt \ \dot
E^{(i)}_{tide}. \eqno(A10)$$ 
We now calculate $\dot E^{(r)}_{tide}$ using the
fourier series 
$$\dot E^{(r)}_{tide} = -\sum_\alpha
\frac{dA_\alpha}{dt} \int d^3x \ U \delta \rho_\alpha Y_{\ell,m}(\theta, \phi)
= \beta \sum_\alpha \frac{dA_\alpha}{dt} Q_\alpha \tilde{f}^*_{2m}.
\eqno(A11)$$  
Substituting equations [A6] \& [A7] into the above equation we obtain
$$\eqalign{\Delta E^{(r)}_{tide} \equiv
\int^{2\pi/\Omega_o}_0 dt \ \dot E^{(r)}_{tide} = \ & \frac{4 \pi
\beta^2}{\Omega^4_o} \sum_{n} Q^2_{n 2 2} \omega^2_{n 2 2} d_{n 2 2} \sum_k 
\frac{(D^{2 2}_k)^2}{[\eta_k(2)]^2} (2 \Omega_* - k)^2 \cr & \quad \quad
+ \frac{\pi
\beta^2}{\Omega^4_o} \sum_{n} Q^2_{n 2 0} \omega^2_{n 2 0} d_{n 2 0} \sum_k 
\frac{(D^{2 0}_k)^2}{[\eta_k(0)]^2} k^2}, \eqno(A12)$$ where the first
term in the above expression is due to the contribution of $m = \pm 2$
modes and the second (which is generally small compared to the first)
is due to the $m = 0$ modes.  Note that the tidal energy as seen by an
observer corotating with the star is positive definite (see Figure 1).
The change in the orbital energy per orbit, $\Delta E_{orb} = -\Delta
E^{(i)}_{tide} = -(\Delta E^{(r)}_{tide} + \Omega_* \Delta L)$, is
given by
$$\eqalign{\Delta E_{orb} = \ & \frac{4 \pi
\beta^2}{\Omega^4_o} \sum_{n} Q^2_{n 2 2} \omega^2_{n 2 2} d_{n 2 2} \sum_k 
\frac{(D^{2 2}_k)^2}{[\eta_k(2)]^2} k (2 \Omega_* - k) \cr & \quad \quad
- \frac{\pi
\beta^2}{\Omega^4_o} \sum_{n} Q^2_{n 2 0} \omega^2_{n 2 0} d_{n 2 0} \sum_k 
\frac{(D^{2 0}_k)^2}{[\eta_k(0)]^2} k^2}. \eqno(A13)$$ 

For the results presented in this paper (e.g., Figures 1 \& 2), we
have used the full fourier series expression for $E_{tide}$, $\Delta
E_{orb}$, etc. (including maintaining the distinction between $C^{2
2}_k$ and $D^{2 2}_k$ which, for simplicity, was dropped following
equation [A7]).  These expressions, however, contain summations over
both mode order and fourier coefficients and are thus somewhat involved.
We now discuss several approximations which allow for
clearer physical understanding of the behavior of and relationship
between $\Delta L$, $\Delta E^{(r)}_{tide}$ and $\Delta E_{orb}$.

The dominant contribution to the
angular momentum and energy deposited comes from the resonant term in
the fourier series, i.e., the integer $k$ for which $\eta_k(m)$ is minimized;
we denote this value by $k_\alpha$. Keeping only this term in the fourier
series, the above expressions for $\Delta L$, $\Delta E^{(r)}_{tide}$
and $\Delta E_{orb}$ simplify to

$$\Delta L \approx -\frac{8 \pi
\beta^2}{\Omega^2_o} \sum_{n} Q^2_{n 2 2} \omega^2_{n 2 2} d_{n 2 2}  
\frac{(D^{2 2}_{k_\alpha})^2}{[\eta_{k_\alpha}(2)]^2} (2 \Omega_* - k_\alpha),
\eqno(A14)$$

$$\eqalign{\Delta E^{(r)}_{tide} \approx \ & \frac{4 \pi
\beta^2}{\Omega^4_o} \sum_{n} Q^2_{n 2 2} \omega^2_{n 2 2} d_{n 2 2} 
\frac{(D^{2 2}_{k_\alpha})^2}{[\eta_{k_\alpha}(2)]^2} (2 \Omega_* - k_\alpha)^2 \cr & \quad \quad
+ \frac{\pi
\beta^2}{\Omega^4_o} \sum_{n} Q^2_{n 2 0} \omega^2_{n 2 0} d_{n 2 0} 
\frac{(D^{2 0}_{k_\alpha})^2}{[\eta_{k_\alpha}(0)]^2} k_{\alpha}^2}, \eqno(A15)$$
and
$$\eqalign{\Delta E_{orb} = - \Delta E^{(i)}_{tide} \approx \ & \frac{4 \pi
\beta^2}{\Omega^4_o} \sum_{n} Q^2_{n 2 2} \omega^2_{n 2 2} d_{n 2 2} 
\frac{(D^{2 2}_{k_\alpha})^2}{[\eta_{k_\alpha}(2)]^2} k_\alpha (2 \Omega_* - 
k_\alpha) \cr & \quad \quad
- \frac{\pi
\beta^2}{\Omega^4_o} \sum_{n} Q^2_{n 2 0} \omega^2_{n 2 0} d_{n 2 0}
\frac{(D^{2 0}_{k_\alpha})^2}{[\eta_{k_\alpha}(0)]^2} k_{\alpha}^2}, 
\eqno(A16)$$ which are the expressions used in the main body of the text.
We note that $\eta_k(0)$ and $\eta_k(2)$ will, of course, in general
be minimized by different $k_\alpha$.  These simplified expressions
are typically accurate to within $\approx 30$ percent in calculating
the energy and angular momentum deposited in the star.  The only
exception is when $\Omega_* \approx \Omega_p$, at which point
equations (A14)-(A16) can underestimate the energy and angular
momentum deposited by up to a factor of $\approx 5$.

As a final simplification, we note that the modes with the most energy
have $|m| = 2$ and $\omega^{(i)}_\alpha \approx 2 \Omega_p$, where
$\omega^{(i)}_\alpha$ is the mode frequency in an inertial reference
frame.  This is because, for $k \gta \Omega_p/\Omega_o$, the $D^{2
2}_k$ are typically larger than the $D^{2 0}_k$ by at least a factor
of a few ($\gta 3$) and so the $m = 0$ mode energies are in general
negligible (with the possible exception of when $\Omega_* \approx
\Omega_p$).  Furthermore, $D^{2 2}_k$ peaks at $k \approx 
2\Omega_p/\Omega_o$ and decreases exponentially for larger $k$.  Thus,
modes with $\omega^{(i)}_\alpha \gta 2 \Omega_p$, while they have
fewer radial nodes and hence larger $Q_\alpha$, have less energy than
modes with $\omega^{(i)}_\alpha \approx 2 \Omega_p$ because of the
rapid decrease of $D^{2 2}_{k_\alpha}$ with $k_\alpha$.  Similarly,
for $\omega^{(i)}_\alpha \lta 2 \Omega_p$, not only is $D^{2
2}_{k_\alpha}$ smaller, but $Q_\alpha$ is also smaller since the modes
have more radial nodes.  	

Thus the dominant contribution to $\Delta L$ and $\Delta E_{orb}$
comes from the $k_\alpha \approx 2 \omp/\Omega_o$ term in the mode
summation, which dramatically simplifies the above expression. From
equation (A14) we see that $\Delta L > 0$ for $\Omega_* \lta \omp$ and
$\Delta L < 0$ for $\Omega_* \gta \omp$, indicating that the tidal
force causes $\Omega_* \rightarrow \ \approx \omp$ (which is called
pseudo-synchronous rotation). From equations [A14]-[A16] it follows
that $\Delta E^{(i)}_{tide} \approx \omp \Delta L $ and $\Delta
E^{(r)}_{tide} \approx (\omp - \Omega_*) \Delta L$.  These relations
show that the energy in the dynamical tide is a frame dependent
quantity. Finally, since $\Delta E_{orb} = -\Delta E^{(i)}_{tide}
\approx - \omp \Delta L$, we see that $\Delta E_{orb} < 0$ for
$\Omega_* \lta \omp$ while $\Delta E_{orb} > 0$ for $\Omega_* \gta
\omp$ (see Figure 2).  The orbital period of a binary system thus 
increases due to the dissipation of the dynamical tide if $\Omega_*
\gta \omp$ since the energy removed from the spin of the star
(which is being slowed down by the tidal torque) exceeds the energy
deposited in modes.

We now show that the dissipation of the equilibrium tide leads to an
analogous relationship among $\dot E^{(i)}_{tide}$, $\dot E^{(r)}_{tide}$, and
$\tau$.  For the equilibrium tide, we find it easiest to explicitly
sum over the $m = \pm 2$ components in $U$, in which case 
 $$U = -V(t) \cos\big[2(\phi -
\phi_{orb}(t) + \Omega_*t)] + V_0(t), \eqno(A17)$$ where $V(t) =
2\beta r^2 Y^*_{2,2}(\pi/2,0)Y_{2,2}(\theta,\phi)/[R(t)/a]^3$ and the
$m = 0$ contribution to the tidal potential is given by $V_0(t) =
\beta r^2 Y^*_{2,0}(\pi/2,0)Y_{2,0}(\theta,\phi)/[R(t)/a]^3$.  The
equilibrium tide represents the hydrostatic response of the primary to
the perturbing gravitational force of the secondary and so the time
dependence of the equilibrium tide follows that of the tidal
gravitational potential.  Thus, the mode amplitude for the equilibrium
tide, as seen by an observer corotating with the star, is given by
(Kumar et al. 1995)
$$A_\alpha = \int d^3x U({\bf r},t) \delta \rho_\alpha.
\eqno(A18)$$  Substituting $U$ and $A_\alpha$ into equations (A4) and (A9)
it is easy to show that, near periastron, $\dot E^{(r)}_{tide} \approx
(\Omega_p -
\Omega_*) \tau_z$ and $\dot E^{(i)}_{tide} \approx \Omega_p \tau_z$,
where we have taken $\dot \phi_{orb}(t) \approx \Omega_p$ and $\dot
R(t) \approx 0$, which are valid near periastron.  For highly eccentric
orbits, the energy in the equilibrium tide peaks strongly near
periastron and so most of the contribution to $\Delta L$ and $\Delta
E_{orb}$ comes from near periastron.  Thus, as with the dynamical
tide, $\Delta E_{orb}
\approx - \omp \Delta L$ for the equilibrium tide.

\vfill\eject
\bigskip
\centerline{\bf Appendix B}
\medskip

Neglecting the rotational modification to the equilibrium structure of the 
star, the perturbed mass and momentum equations, for a rotating star,
in the Cowling approximation are 
$$\delta \rho + \nabla\cdot(\rho\bxi) = 0
\eqno(B1)$$ and $$-\omega^2 \rho \bxi = -{\bf \nabla}\delta p + \delta
\rho\, {\bf g} + 2 \rho i \omega \bxi \times {\bf \Omega_*}, \eqno(B2)$$
where $\delta$ denotes an Eulerian perturbation and we
have assumed a time dependence of $\exp(i\omega_\alpha t)$.  The
Centrifugal force does not appear in equation (B2) because it has no
Eulerian perturbation.\footnote{$^{12}$}{Equivalently, the Centrifugal
force evaluated at the perturbed position of the fluid is identically
canceled by the equilibrium structure of the star evaluated at the
perturbed position.} 
The components of the momentum equation for adiabatic oscillations can
be written as

$$-\omega^2 \rho \xi_r = -\frac{d\delta
p}{dr} - \frac{g \delta p}{c^2_s} - N^2 \xi_r \rho +
2i\omega\rho\Omega_*\xi_\phi \sin\theta,
\eqno(B3)$$ $$-\omega^2 \rho \xi_\theta = -\frac{1}{r}\frac{d\delta
p}{d\theta} + 2i\omega\rho\Omega_*\xi_\phi \cos\theta, \eqno(B4)$$ and
$$-\omega^2 \rho \xi_\phi = -\frac{1}{r\sin\theta}\frac{d\delta
p}{d\phi} - 2i\omega\rho\Omega_* \big(\xi_r \sin\theta + \xi_\theta
\cos\theta\big) , \eqno(B5)$$
where we have taken ${\bf \Omega_*}=\Omega_* {\bf \hat z}$.
In the region of wave propagation $\xi_\phi \sim N\xi_r/\omega$, and thus the
buoyancy term in the radial momentum equation is larger than the
Coriolis term by a factor of $N/\Omega_*$; we thus neglect the latter.
Away from the equator, $\xi_r \sin\theta \sim (\xi_\theta \cos\theta) 
\omega/N$ and thus we neglect the radial displacement in the $\phi$
component of the momentum equation.  These approximations constitute
the `traditional approximation' (Bildsten et al. 1996; Chapman \&
Lindzen 1970; Unno et al. 1989) and are only valid when $\omega_\alpha
\ll N$ and $\Omega_* \ll N$.

Solving for $\xi_\theta$ and $\xi_\phi$ in terms of $\delta p$ and
taking the perturbed variables to be proportional to $\exp(i m \phi)$
we obtain the following coupled radial equations
$$\frac{1}{r^2}\frac{d}{dr}(r^2 \xi_r) + \frac{\delta p}{\rho c^2_s}\left[
   1 - {\lambda c_s^2\over r^2\omega^2}\right] -
\frac{g\xi_r}{c^2_s} = 0,\eqno(B6)$$
and $$\frac{d \delta p}{dr} +
\frac{g \delta p}{c^2_s} - \rho(\omega^2 - N^2)\xi_r = 0, \eqno(B7)$$
where the angular eigenfunctions 
(called the Hough functions) and eigenvalues ($\lambda$) are the
solutions to the following eigenvalue problem $$L_q H = - \lambda H,
\eqno(B8)$$ where 
$$L_q = \frac{-m^2}{(1-x^2)(1-q^2x^2)} +
\frac{d}{dx}\left[\frac{1-x^2}{1-q^2x^2}\frac{d}{dx}\right] - 
qm\frac{1+q^2x^2}{\big(1-q^2x^2\big)^2}, \eqno(B9)$$ where $x =
\cos\theta$ and where the angular operator (and thus $\lambda$ and the Hough
functions) depends on the stellar rotation rate and mode frequency
through the parameter $q \equiv 2\Omega_*/\omega_\alpha$.  We note
that, for a fixed $|m|$, $|\omega_\alpha|$, and $|\Omega_*|$, it
is the sign of $m\Omega_*/\omega_\alpha$ that determines the effect
of the Coriolis force.

Figure (5a) shows $\lambda$ as a function of $q\equiv
2\Omega_*/\omega_\alpha$ for the even Hough function with the smallest
value of $\lambda$, which is a generalization of $\ell=2$.
For $m = -2$, $\lambda \rightarrow m^2 = 4$ for $q \gg 1$ and thus the mode
structure of the $m = -2$ modes is only weakly modified by the
rotation of the star.  More detailed discussion of the Hough functions
is given in the above references.  In what follows, we focus on the
aspects of the modified mode structure relevant to tidal excitation.

The overlap integral for a mode can be written as $$Q_\alpha \equiv
Q_{\alpha, r} Q_{angle} = \int dr\, r^4 \delta \rho_{\alpha}(r) \int
d\Omega\, H^q_{\lambda m}(\theta) Y_{2m}(\theta, \phi) e^{-im\phi}.
\eqno(B10)$$   It can easily be shown using a WKB analysis, or 
verified numerically, that $Q_{\alpha, r}$, the radial contribution to
the overlap integral, is only a function of the number of radial nodes
of the mode, {\it i.e.}, for a fixed $n$, $Q_{\alpha, r}$ is
independent of $\Omega_*$, $\omega_\alpha$, $\lambda$ and $m$. The
overlap of the even Hough function with the smallest value of
$\lambda$ (the generalization of $\ell = 2$) with the quadrupole
spherical harmonic, $Q_{angle}$, is shown in Figure (5b) as a function
of $q$. For $q \lta 1$, the associated Legendre polynomials are very
good approximations to the Hough functions, but for $q \gta 1$ the
Hough functions are concentrated near the equator.

Using the information in Figure (5), there is a simple prescription
for calculating g-mode frequencies and overlap integrals of a rotating
star in the `traditional approximation'. To calculate the frequency of
a quadrupole g$_n$ mode, take the value of $q_\alpha =
2\Omega_*/\omega_n^0$ (superscript $o$ denotes the value for
non-rotating star) and determine $\lambda$ from Figure (5a). The
frequency in the rotating star is, to first approximation, given by
$\omega_n^0 \lambda^{1/2}/\sqrt{6}$. This frequency can be used to
re-evaluate $q_\alpha$ and in turn $\lambda$ and $\omega_\alpha$, and
this process can be iterated to the desired accuracy.  The angular
overlap integral, $Q_{angle}$, can be determined from Figure (5b)
using the resulting value of $q_\alpha$ and, since $Q_{\alpha,r}$ is a
function of $n$ alone, independent of $\Omega_*$, we obtain $Q_\alpha$
for the g-modes of the rotating star.

\vfill\eject

\centerline{\bf REFERENCES}
\bigskip

\refind Bell, J.F., Bessell, M.S., Stappers, B.W., Bailes, M., Kaspi, V.M. 
1995, ApJ, 447, L117

\ni Bildsten, L., Ushomirsky, G., and Cutler, C. 1996, ApJ, 460, 827

\ni Chapman, S. and Lindzen, R. S. 1970, {\it Atmospheric Tides} 
    (Dordrecht: Reidel)

\ni Goldreich, P. and Nicholson, P. D. 1989, ApJ, 342, 1079

\ni Hut, P. 1981, A \& A, 99, 126

\refind Johnston, S., Manchester, R.N., Lyne, A.G., Bailes, M., Kaspi, V.M.,
    Guojun, Q., and D'Amico, N., 1992, ApJ 387, L37

\refind Kaspi, V.M., Baile, M., Manchester, R.N., Stappers, B.W., and Bell, 
  J.F. 1996, Nature, 381, 584

\refind Kaspi, V.M., Johnston, S., Bell, J.F., Manchester, R.N., Bailes, M.,
    Bessell, M., Lyne, A.G., and D'amico, N., 1994, ApJ, 423, L43

\ni Kumar, P. 1994, ApJ, 428, 827

\ni Kumar, P., Ao, C. O., and Quataert, E. J.,1995, ApJ, 449, 294

\ni Kumar, P. and Goodman, J., 1996, 466, 946

\ni Kumar, P. and Quataert, E. J., 1996,  

\ni Lai, D. 1996, ApJ, 466, L35

\ni Lai, D., Bildsten, L., and Kaspi, V.M 1995, ApJ, 452, 819

\refind McConnell, D., McCulloch, P.M., Hamilton, P.A., Ables, J.G., Hall, P.J.,
Jacka, C.E., and Hunt, A.J., 1991, MNRAS 249, 654

\ni Press, W.H., and Teukolsky, S.A., 1977, ApJ, 213, 183

\ni Quataert, E.J., Kumar, P., and Ao, C.O. 1996, ApJ, 463, 284

\ni Saio, H. and Cox, J.P. 1980, ApJ, 236, 549

\refind Unno, W., Osaki, Y., Ando, H., Saio, H., and Shibahashi, H.,
    1989, {\it Nonradial Oscillations of Stars}, 2nd Ed.  (University
of Tokyo Press)

\vfill\eject
\centerline{}
\vskip 2.0truecm

\centerline{\bf Table 1}
\smallskip
\centerline{Parameters for the SMC binary system}
\medskip
$$
\vbox{\offinterlineskip
\hrule
\halign{&\vrule#\hfil &
    \strut\quad\hfil#\hfil\quad\cr
  \noalign{\hrule}
  height2pt&\omit&&\omit&\cr
  & Orbital period ($P_{orb}$)  && \hfill $4.421\times 10^6$ s (51.17 days)~~~ &\cr
  & e && \hfill 0.8079 ~~~~ &\cr
  & $P_{orb}/\dot P_{orb}$ && \hfill $-4.63\times 10^5$ yr.~~~ &\cr
  & 1/$|\dot e|$ && $\hfill >1.5\times 10^5$ yr. ~~~&\cr
  & $\Omega_p/2\pi$ && \hfill 3.61 $\mu$Hz ~~~&\cr
  & $M_*$  && \hfill $\approx$ 8.8 $M_\odot$ ~~~&\cr
  & $R_*$  && \hfill $\approx$ 6.0 $R_\odot$ ~~~&\cr
  & $\Omega_*/2\pi$ && \hfill 10.1 $\mu$Hz for $\hat\Omega_*=0.5$ ~~~&\cr
  height2pt&\omit&&\omit&\cr}
 \hrule}
$$

\vfill\eject
\centerline{}
\vskip 2.0truecm

\centerline{\bf Table 2}
\smallskip
\centerline{Frequencies and radiative dissipation times (energy) for the}
\centerline{low order quadrupole g-modes of a non-rotating B-star} 
\centerline{of mass 8.8 $M_\odot$ and radius 6.0 $R_\odot$}
\medskip

$$
\vbox{\offinterlineskip
\hrule
\halign{&\vrule#\hfil &
    \strut\quad\hfil#\hfil\quad\cr
   height2pt&\omit&&\omit&&\omit&\cr
   & n (radial nodes) \hfil&& frequency ($\mu$Hz) && $\Gamma_\alpha^{-1}$ (yrs)&\cr
   height2pt&\omit&&\omit&&\omit&\cr
  \noalign{\hrule}
  height2pt&\omit&&\omit&\cr
  & 1 && 34.66 && 8.7x10$^3$&\cr
  & 2 && 23.04 && 2.5x10$^3$&\cr
  & 3 && 17.17 && 1.1x10$^3$&\cr
  & 4 && 12.58 && 4.6x10$^2$&\cr
  & 5 && 11.21 && 1.6x10$^2$&\cr
  & 6 &&  9.58 && 47.9 &\cr
  & 7 &&  8.43 && 14.5 &\cr
  & 8 &&  7.56 && 5.6  &\cr
  & 9 &&  6.83 && 2.5  &\cr
  & 10&&  6.23 && 1.1  &\cr
  height2pt&\omit&&\omit&&\omit&\cr}
 \hrule}
$$

\vfill\eject
\centerline{\bf Figure Captions}
\bigskip

\ni FIG. 1.--- (a) The energy in the dynamical tide for the B-star of the 
SMC Binary, as seen by an observer corotating with the B-star
($E^{(r)}_{tide}$), as a function of the B-star's rotation rate (in
units of $\sqrt{GM_*/R_*^3}$).  (b) The energy in the dynamical tide for
the B-star of the SMC Binary, as seen by an inertial observer
($E^{(i)}_{tide}$). Note that $E^{(r)}_{tide} > 0$ for all rotation rates while
$E^{(i)}_{tide} > 0$ for $\Omega_* \lta \Omega_p$, i.e., $\os \lta
0.2$ (solid line) but $E^{(i)}_{tide} < 0$ for $\Omega_* \gta
\omp$ (dashed line). All relevant $\ell=2$ g-modes of $m=0, \pm2$ were
included in the calculations.  These calculations (as well as those
of Fig. 2) were made using the full fourier series expressions for
$E^{(i)}_{tide}$ and $E^{(r)}_{tide}$ given in Appendix A.

\ni FIG. 2.--- (a) The change in the orbital energy per orbit, 
$\Delta E_{orb}$, and (b) the resulting orbital evolution timescale,
$P_{orb}/\dot P_{orb}$, for the radiative dissipation of the dynamical
tide in a rigidly rotating B-star of the PSR J0045-7319 Binary as a
function of the B-star's rotation rate (in units of $\sqrt{GM_*/R_*^3}$);
the calculation included all quadrupole g-modes of $m=\pm2$ as well as $m=0$.
For $\Omega_* \lta \Omega_p$, i.e., $\os \lta 0.2$, $\Delta
E_{orb}$ and $P_{orb}/\dot P_{orb}$ are negative (solid line) while
for $\Omega_* \gta \Omega_p$, $\Delta E_{orb}$ and $P_{orb}/\dot
P_{orb}$ are positive (dashed line); please see the text for a physical
explanation of this result.

\ni FIG. 3.--- The local radiative dissipation rate for a quadrupole
g-mode of frequency $\approx 17 \mu$Hz for the B-star of the SMC
Binary; this is the mode predominantly excited for rapid retrograde
rotation of the B-star ($\os=-0.3$). The dissipation peaks near the outer 
turning point of the mode, which occurs at $r \approx 0.85 R_*$, and so the
angular momentum and energy of the mode are deposited well beneath the
surface of the star.

\ni FIG. 4.--- The adiabatic energy flux in a g-mode of the B-star in the 
SMC Binary for rigidly (dashed line) and differentially rotating
(solid line) stars which have the same rotation frequency in the
interior.  The differential rotation is centered at $r \approx 0.8
R_*$, occurs over a width of $\approx 0.1 R_*$, and is such that the
minimum wave frequency (which is $\approx 17 \mu$Hz in the core) is $\approx 1
\mu$Hz at $r \approx 0.8 R_*$.  The small wavelength in the
differentially rotating layer implies that the mode dissipation is
strongly enhanced.  Note that energy flux is nearly constant across the 
differentially rotating layer, or in other words the differential rotation
does not significantly reflect the wave.

\ni FIG. 5.--- (a) The smallest value of the angular eigenvalue $\lambda$ as
a function of $q \equiv 2\Omega_*/\omega_\alpha$ for even parity modes
(these are the generalization of the $\ell = 2$ modes for a rotating
star and so $\lambda \rightarrow \ell(\ell+1) = 6$ as $q \rightarrow
0$).  (b) $Q_{angle}$, the overlap of the angular eigenfunction in the
rotating star with the quadrupole spherical harmonic, for the same
modes as part (a).  Since the radial contribution to the overlap
integral is, for a fixed number of radial nodes, independent of
$\Omega_*$, $Q_{angle}$ give the variation of the overlap integral
with the stellar rotation rate.

\bye

%% file: macro.tex

\def\chaphead{}
\def\ni{\noindent}

\font\tfont=cmbxti10
\font\eightrm=cmr8
\font\eightit=cmti8
\font\sixrm=cmr6
\font\eightmit=cmmi8
\font\sixmit=cmmi6
\def\absmath{\textfont0=\eightrm \scriptfont0=\sixrm
	      \textfont1=\eightmit \scriptfont1=\sixmit}
\def\absfont{\let\rm=\eightrm \let\it=\eightit \rm\absmath}
\font\twelverm=cmr12
\font\twelveit=cmti12
\font\tenrm=cmr10
\font\twelvemit=cmmi12
\font\tenmit=cmmi10
\def\regmath{\textfont0=\twelverm \scriptfont0=\tenrm
	      \textfont1=\twelvemit \scriptfont1=\tenmit}
\def\peterfont{\let\rm=\twelverm \let\it=\twelveit \rm\regmath}
%
%

\newfam\vecfam

\textfont\vecfam=\tfont \scriptfont\vecfam=\seveni
\scriptscriptfont\vecfam=\fivei


\def\spose#1{\hbox to 0pt{#1\hss}}

\font\eightrm=cmr8

\def\s{\ifmmode \widetilde \else \~\fi} 
     
\def\section{\S}
\newcount\notenumber
\notenumber=1
\newcount\eqnumber
\eqnumber=1
\newcount\fignumber
\fignumber=1
\newbox\abstr


\def\s{{\rm\,s}}

\def\note#1{\footnote{$^{\the\notenumber}$}{#1}\global\advance\notenumber by 1}
\def\foot#1{\raise3pt\hbox{\eightrm \the\notenumber}
     \hfil\par\vskip3pt\hrule\vskip6pt
     \noindent\raise3pt\hbox{\eightrm \the\notenumber}
     #1\par\vskip6pt\hrule\vskip3pt\noindent\global\advance\notenumber by 1}

\def\abstract#1{\setbox\abstr=\vbox{\hsize 5.0truein{\par\noindent#1}}
    \centerline{ABSTRACT} \vskip12pt \hbox to \hsize{\hfill\box\abstr\hfill}}
     
\def\Dt{\spose{\raise 1.5ex\hbox{\hskip3pt$\mathchar"201$}}}    
\def\dt{\spose{\raise 1.0ex\hbox{\hskip2pt$\mathchar"201$}}}    

\def\new{{\rm\chaphead\the\eqnumber}\global\advance\eqnumber by 1}
\def\ref#1{\advance\eqnumber by -#1 \chaphead\the\eqnumber
     \advance\eqnumber by #1 }
\def\last{\advance\eqnumber by -1 {\rm\chaphead\the\eqnumber}\advance
     \eqnumber by 1}
\def\eqnam#1{\xdef#1{\chaphead\the\eqnumber}}
     
\def\nfig{\chaphead\the\fignumber\global\advance\fignumber by 1}
\def\nfiga#1{\chaphead\the\fignumber{#1}\global\advance\fignumber by 1}
\def\rfig#1{\advance\fignumber by -#1 \chaphead\the\fignumber
     \advance\fignumber by #1}
\def\fignam#1{\xdef#1{\chaphead\the\fignumber}}

\def\lta{\mathrel{\spose{\lower 3pt\hbox{$\mathchar"218$}}
     \raise 2.0pt\hbox{$\mathchar"13C$}}}
\def\gta{\mathrel{\spose{\lower 3pt\hbox{$\mathchar"218$}}
     \raise 2.0pt\hbox{$\mathchar"13E$}}}
     

\magnification=\magstep1
\parskip=3pt


\font\gkvec=cmmib10                         

\def\bxi{\hbox{{\gkvec\char24}}}           

     
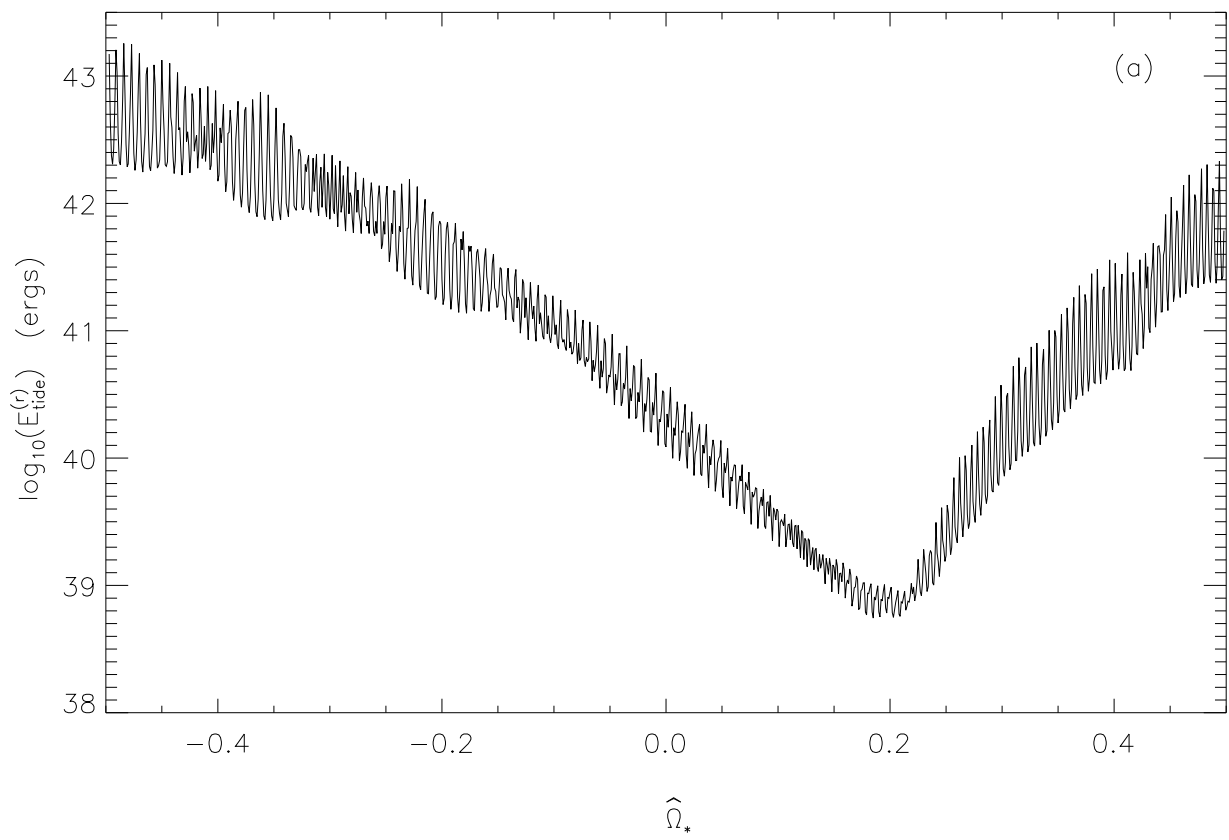
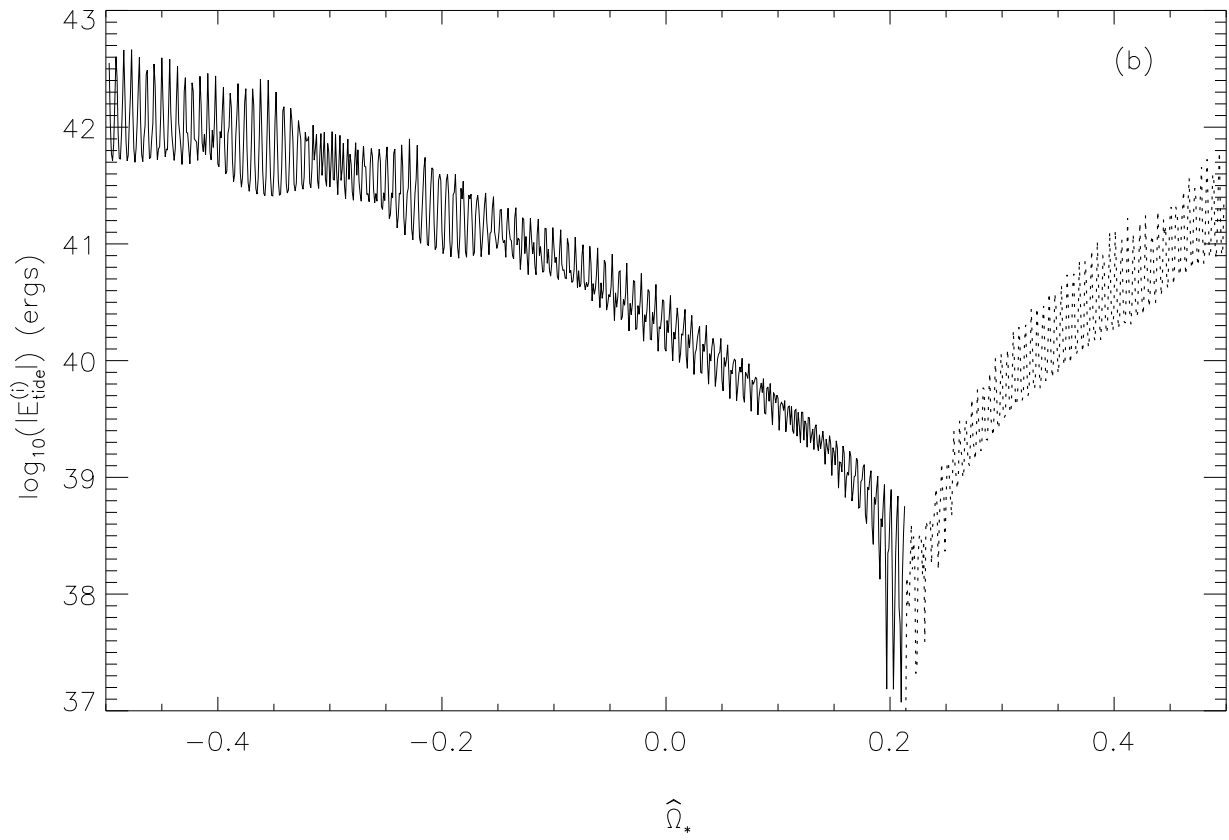
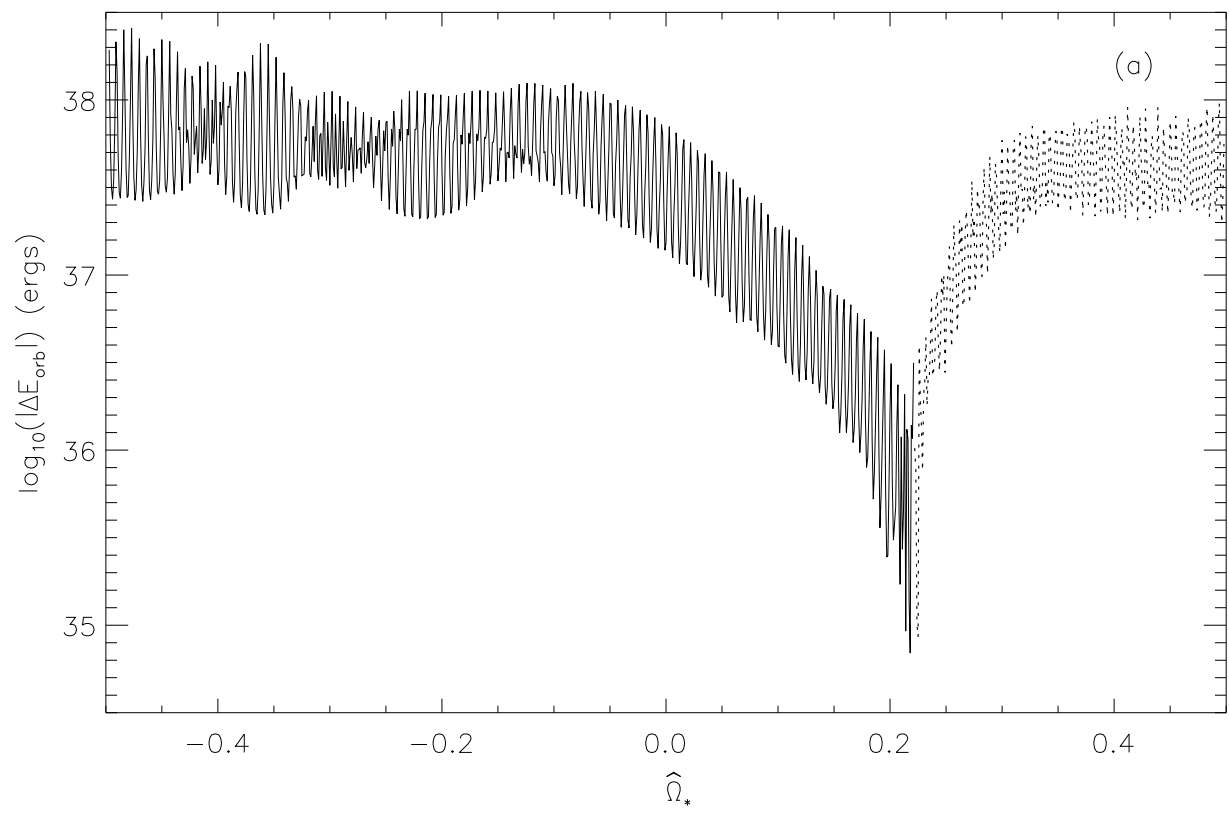
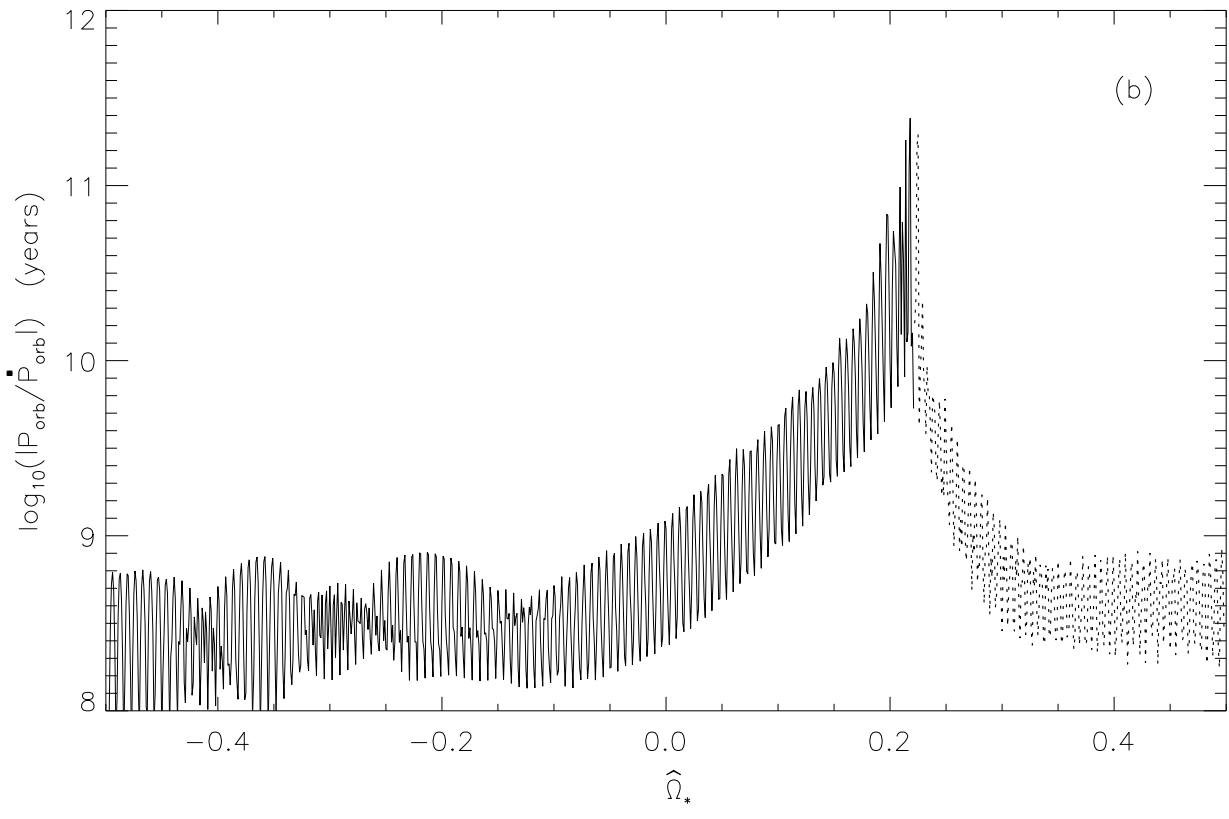
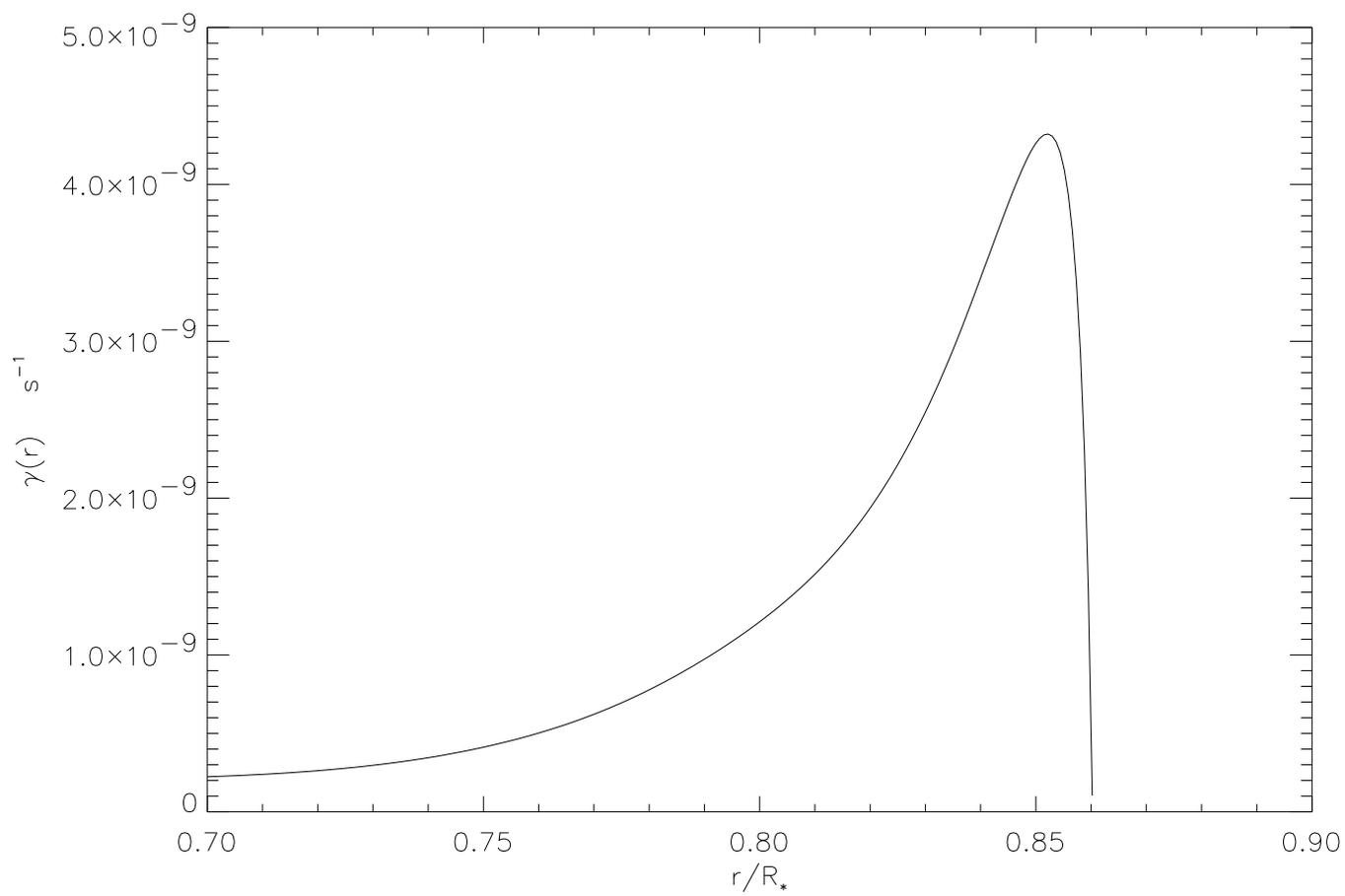
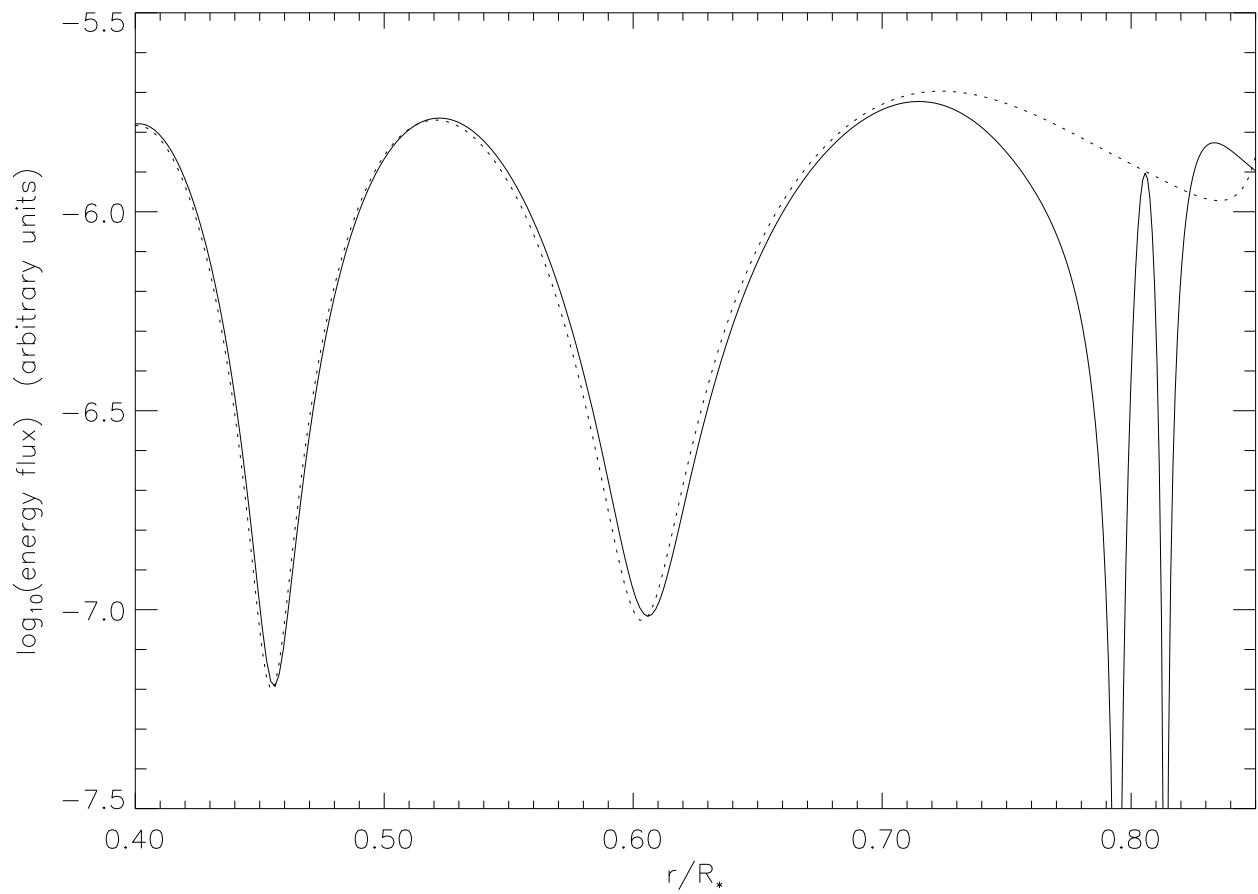
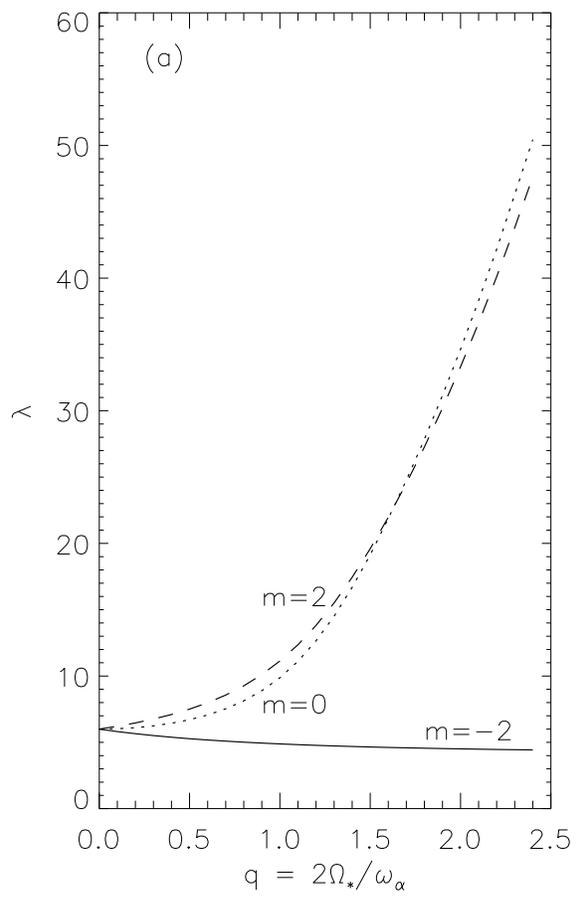
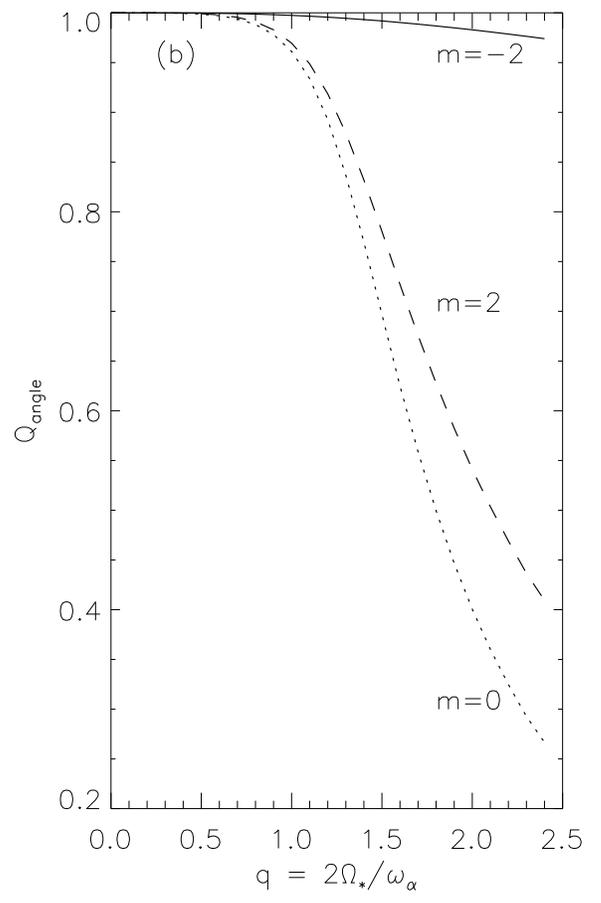
